\documentclass[traditabstract]{aa} 
\usepackage[T1]{fontenc}
\usepackage[latin1]{inputenc}
\usepackage{longtable}
\usepackage{supertabular}
\usepackage{graphicx}
\usepackage{array}
\usepackage{tabularx}
\usepackage{ragged2e}
\usepackage{color}
\usepackage{txfonts}
\usepackage{natbib}
\usepackage{here}
\usepackage{multirow}
\usepackage{wasysym}
\begin{document}

   \title{Forbidden oxygen lines in comets at various heliocentric distances\thanks{Based on observations made with ESO Telescope at the La Silla Paranal Observatory under programs ID 268.C-5570, 270.C-5043, 073.C-0525, 274.C-5015, 075.C-0355, 080.C-0615, 280.C-5053, 086.C-0958 and 087.C-0929.}}
   


\newcommand*\samethanks[1][\value{footnote}]{\footnotemark[#1]}
   \author{A. Decock
          \inst{1}
			,
          E. Jehin\inst{1}			,
			D. Hutsemékers\inst{1}
			,
			J. Manfroid\inst{1}
          }

   \institute{Institut d'Astrophysique et Géophysique et Océanographie, Université de Liège, Allée du 6 août 17, 4000 Liège, Belgium\\
   	              \email{adecock@ulg.ac.be}
             }

   \date{Received Month, year; accepted Month, year}

  \abstract
     {We present a study of the three forbidden oxygen lines [OI] located in the optical region (i.e., $5577.339~\r{A}$ (the green line), $6300.304~\r{A}$ and $6363.776~\r{A}$ (the two red lines)) in order to better understand the production of these atoms in cometary atmospheres. The analysis is based on 48 high-resolution and high signal-to-noise spectra collected with UVES at the ESO VLT between 2003 and 2011 referring to 12 comets of different origins observed at various heliocentric distances.
  The flux ratio of the green line to the sum of the two red lines is evaluated to determine the parent species of the oxygen atoms by comparison with theoretical models. This analysis confirms that, at about 1~AU, H$_{2}$O is the main parent molecule producing oxygen atoms. At heliocentric distances > 2.5~AU, this ratio is changing rapidly, an indication that other molecules are starting to contribute. CO and CO$_{2}$, the most abundant species after H$_{2}$O in the coma, are good candidates and the ratio is used to estimate their abundances. We found that the CO$_{2}$ abundance relative to H$_{2}$O in comet C/2001 Q4 (NEAT) observed at 4~AU can be as high as $\sim$70$\%$.
The intrinsic widths of the oxygen lines were also measured. The green line is on average about 1~km~s$^{-1}$ broader than the red lines while the theory predicts the red lines to be broader. This might be due to the nature of the excitation source and/or a contribution of CO$_{2}$ as parent molecule of the $5577.339~\r{A}$ line. At 4~AU,  we found that the width of the green and red lines in comet C/2001 Q4 are the same which could be explained if CO$_{2}$ becomes the main contributor for the three [OI] lines at high heliocentric distances.}


\authorrunning{Decock et al.}
\titlerunning{[OI] lines in comets at various heliocentric distances}

   \maketitle


\section{Introduction}

   Comets are small bodies formed at the birth of the Solar system 4.6 billion years ago. Since they did not evolve much, they are potential witnesses of the physical and chemical processes at play at the beginning of our Solar system \citep{Ehrenfreund2000}. Their status of "fossils" gives them a unique role to understand the origins of the Solar system, not only from the physical and dynamical point of view but also from the chemical point of view (thanks to the knowledge of the compounds of the nucleus). When the comet gets closer to the Sun, the ices of the nucleus sublimate to form the coma (the comet atmosphere) where oxygen atoms are detected. Oxygen is an important element in the chemistry of the Solar system given its abundance and its presence in many molecules including H$_{2}$O which constitutes 80$\%$ of the cometary ices. Oxygen atoms are produced by the photo-dissociation of molecules coming from the sublimation of the cometary ices. Chemical reactions Eq.(\ref{eqchimie}) to (\ref{eqchimiefin}) involve possible parent molecules \citep{Bhardwaj2012, Festou1981}.
   
\begin{eqnarray}
\label{eqchimie}
\mathrm{H_{2}O + h \nu} &\rightarrow& \mathrm{H_{2} + O^{*}(^{1}D)} \\
\mathrm{H_{2}O + h \nu} &\rightarrow& \mathrm{H_{2} + O^{*}(^{1}S)} 
\end{eqnarray}
\begin{eqnarray}
\mathrm{CO_{2} + h \nu} &\rightarrow& \mathrm{CO + O^{*}(^{1}D)} \\
\mathrm{CO_{2} + h \nu} &\rightarrow& \mathrm{CO + O^{*}(^{1}S)}
\end{eqnarray}
\begin{eqnarray}
\mathrm{CO + h \nu} &\rightarrow& \mathrm{C + O^{*}(^{1}D)} \\
\mathrm{CO + h \nu} &\rightarrow& \mathrm{C + O^{*}(^{1}S)}
\label{eqchimiefin}
\end{eqnarray}

\begin{figure}[h!]
\includegraphics[width=\columnwidth]{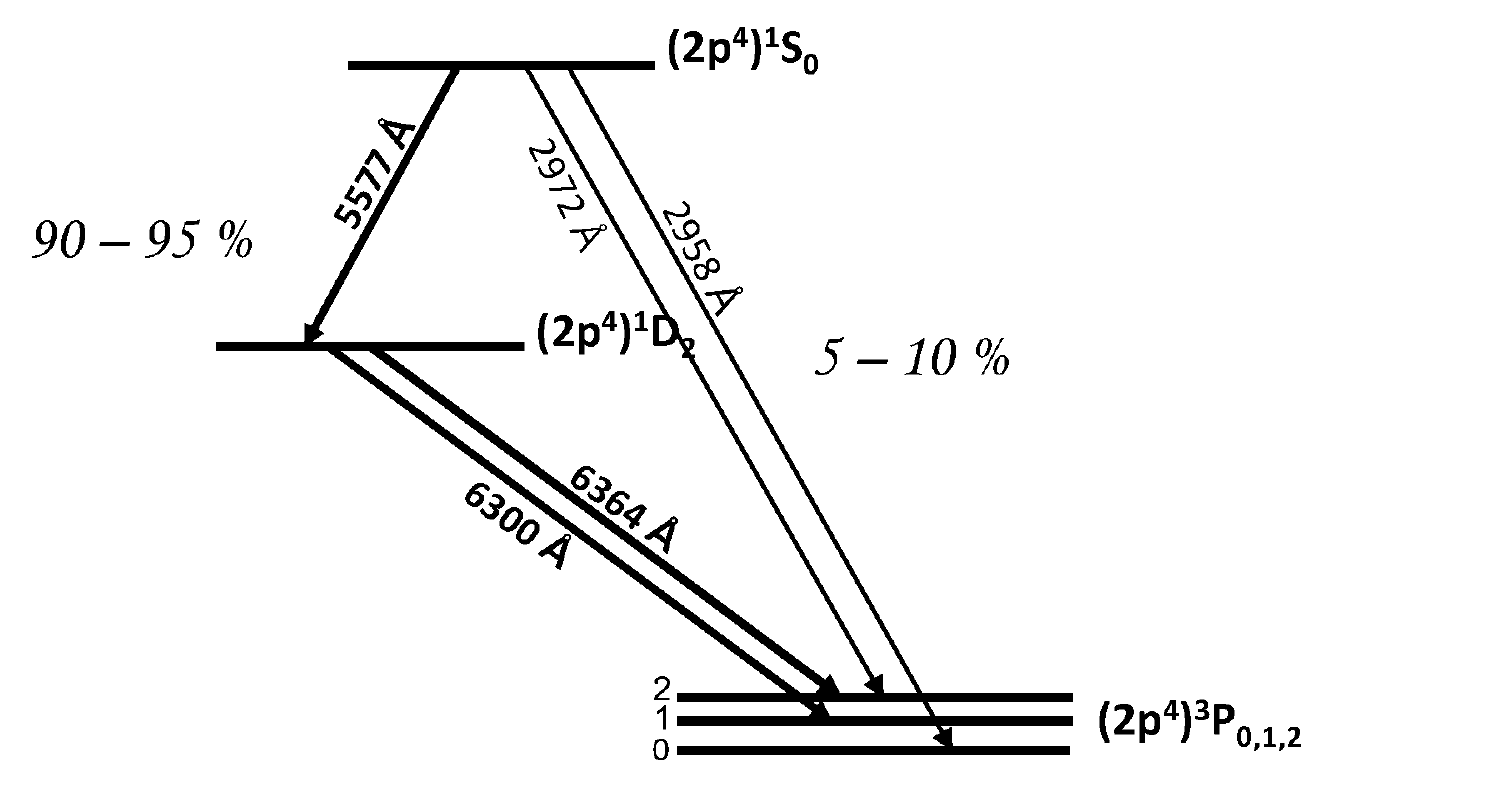}
\caption{Energy level diagram for [OI] lines.}
\label{transition}
\end{figure}

  Oxygen atoms have been detected in comets through the three forbidden lines observed in emission at 5577.339~\r{A} (the green line), 6300.304 \r{A} and 6363.776 \r{A} (the red lines) \citep{Swings1962}. These lines are coming from the deexcitation of the upper state (2p$^{4}$)$^{1}$S$_{0}$ to the (2p$^{4}$)$^{1}$D$_{2}$ state for the 5577.339 $\AA$ line and from the (2p$^{4}$)$^{1}$D$_{2}$ state to the (2p$^{4}$)$^{3}$P$_{1,2}$ states for the doublet red lines (see Fig.~\ref{transition}). The lifetime of the oxygen in the $^{1}$D state is $\sim$ 110 ~s which is much longer than the lifetime of $\sim$1~s for the O($^{1}$S). The measurement of the green line is more difficult owing to its fainter intensity and the many C$_{2}$ lines located around it. One of the first theoretical studies was carried out by \cite{Festou1981}. They reviewed the production rate of O($^{1}$S) and O($^{1}$D) obtained in the laboratory from H$_{2}$O, CO and CO$_{2}$ photo-dissociations and measured the corresponding $^{1}$S/$^{1}$D ratio. Recently, \cite{Bhardwaj2012} made a new model for oxygen atom emissions  and calculated the production rate of O($^{1}$S) and O(${^1}$D) for chemical reactions (\ref{eqchimie}) to (\ref{eqchimiefin}). The estimated $^{1}$S/$^{1}$D ratios are significantly different from \cite{Festou1981} and are given in Table~\ref{bhardwaj}.
  
\begin{table}
\centering
\begin{tabular}{l || c c || c | r}
\hline
\hline
Parents & \multicolumn{2}{|c||}{Emission rate (s$^{-1}$)} & \multicolumn{2}{c}{Ratio}  \\
 & O($^{1}$S) & O($^{1}$D) & $^{1}$S/$^{1}$D & $^{1}$S/$^{1}$D$^{a}$  \\
\hline
H$_{2}$O & 6.4~10$^{-8}$ & 8.0~10$^{-7}$ & 0.080 & $\sim$0.1  \\
CO & 4.0~10$^{-8b}$ & 5.1~10$^{-8}$ & 0.784 & $\sim$1   \\
CO$_{2}$ & 7.2~10$^{-7}$ & 1.2~10$^{-6}$ & 0.600 & $\sim$1  \\
\hline
\end{tabular}
\caption{\label{bhardwaj}Emission rates and $^{1}$S/$^{1}$D ratios relevant to the quiet Sun obtained by \cite{Bhardwaj2012} for photo-dissociation reactions at 1~AU. \newline
$^{a}$ Ratios obtained by \cite{Festou1981} are given in the last column for comparison. \newline
 $^{b}$ This rate comes from \cite{Huebner1979}.\newline
 }
\end{table}

\paragraph{}

Up to now, little systematic researchwork has been done to study these lines at various heliocentric distances because the detection of the forbidden lines requires both high spectral and high spatial resolutions. \cite{Morrison1997} observed oxygen in comet C/1996 B2 (Hyakutake) with the 1-m Ritter Observatory telescope,  \cite{Cochran2001} and \cite{Cochran2008} analysed [OI] lines in the spectra of 8 comets observed at the McDonald Observatory (see \cite{Bhardwaj2012} for a complete review of these measurements).The present paper reports the results obtained for a homogeneous set of high quality spectra of 12 comets of various origins observed since 2003 with the UVES spectrograph mounted on the 8-m Kueyen telescope of the ESO VLT.\\
  First, we measured the intensities of the three forbidden oxygen lines. Two ratios were evaluated : the ratio between the two red lines, $I_{6300}/I_{6364}$, and the ratio of the green line to the sum of the red lines, $I_{5577}/(I_{6300}+I_{6364})$ which we shall denote as G/R hereafter. The purpose of the latter is to determine the main parent molecule of the oxygen atoms by comparing our results with the \cite{Bhardwaj2012} effective excitation rates. 
We also followed how the G/R ratio depends on the heliocentric distance by analysing spectra of comets C/2001 Q4 (NEAT) and C/2009 P1 (Garradd) at small and large distances from the Sun. \\
Finally, we measured the Full Width at Half Maximum (FWHM) of the three lines. There is a long-standing debate on the FWHM measurement because the first observations made by \cite{Cochran2008} are in contradiction with the theory. The intrinsic width of the green line is wider than the red ones while the theory predicts the opposite \citep{Festou1981a}. 
\paragraph{}
In this paper, we present the analysis of the three forbidden oxygen lines in various comets at different heliocentric distances. The observation of all the comets of our sample are explained in section 2. In sections 3 and 4, details about data reduction and data analysis are described. Results of the two ratios and the FWHM of the [OI] lines are given in section 5 followed by discussion in section 6.

\section{Observations}

\begin{figure}[h!]
\includegraphics[width=\columnwidth]{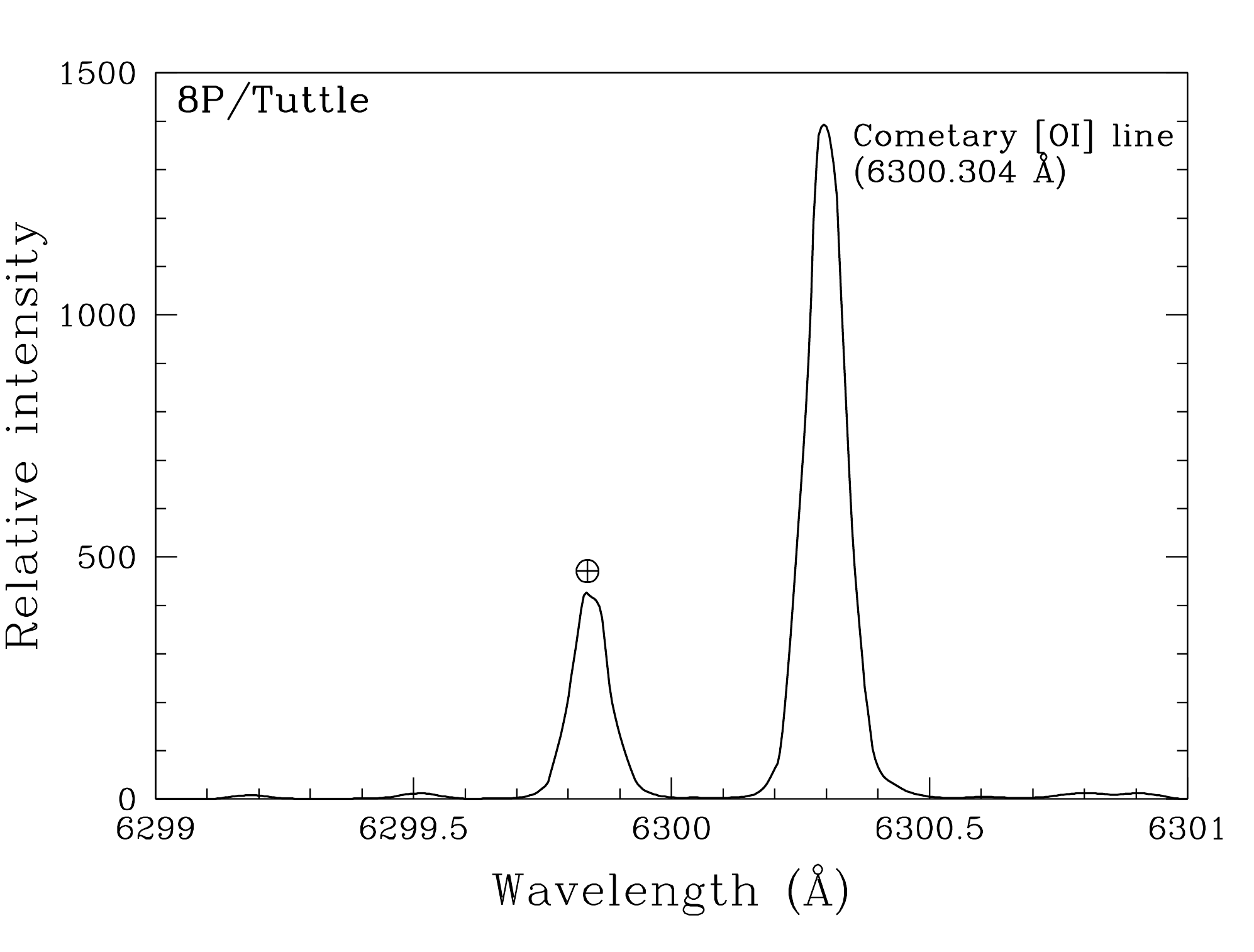}
\includegraphics[width=\columnwidth]{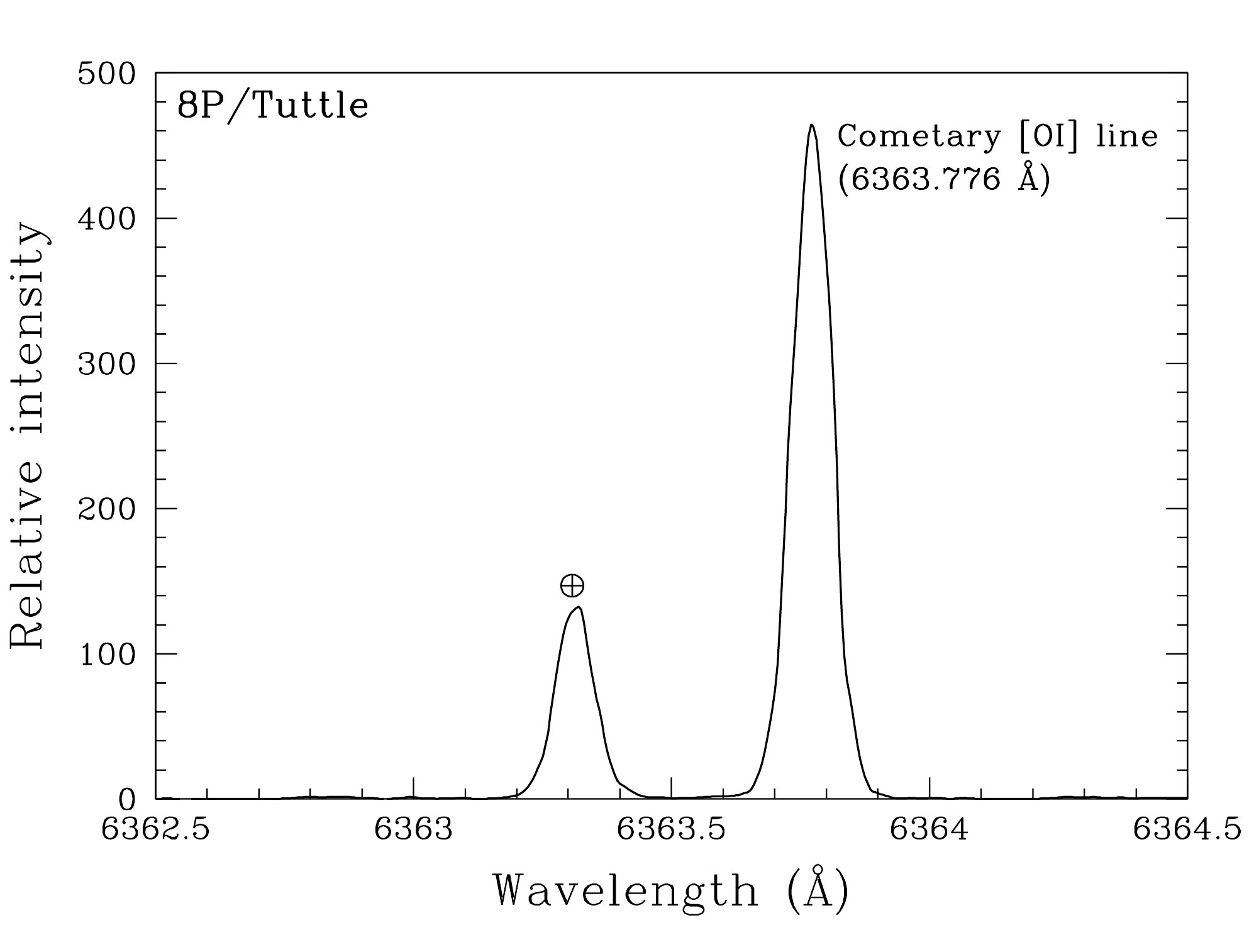}
\includegraphics[width=\columnwidth]{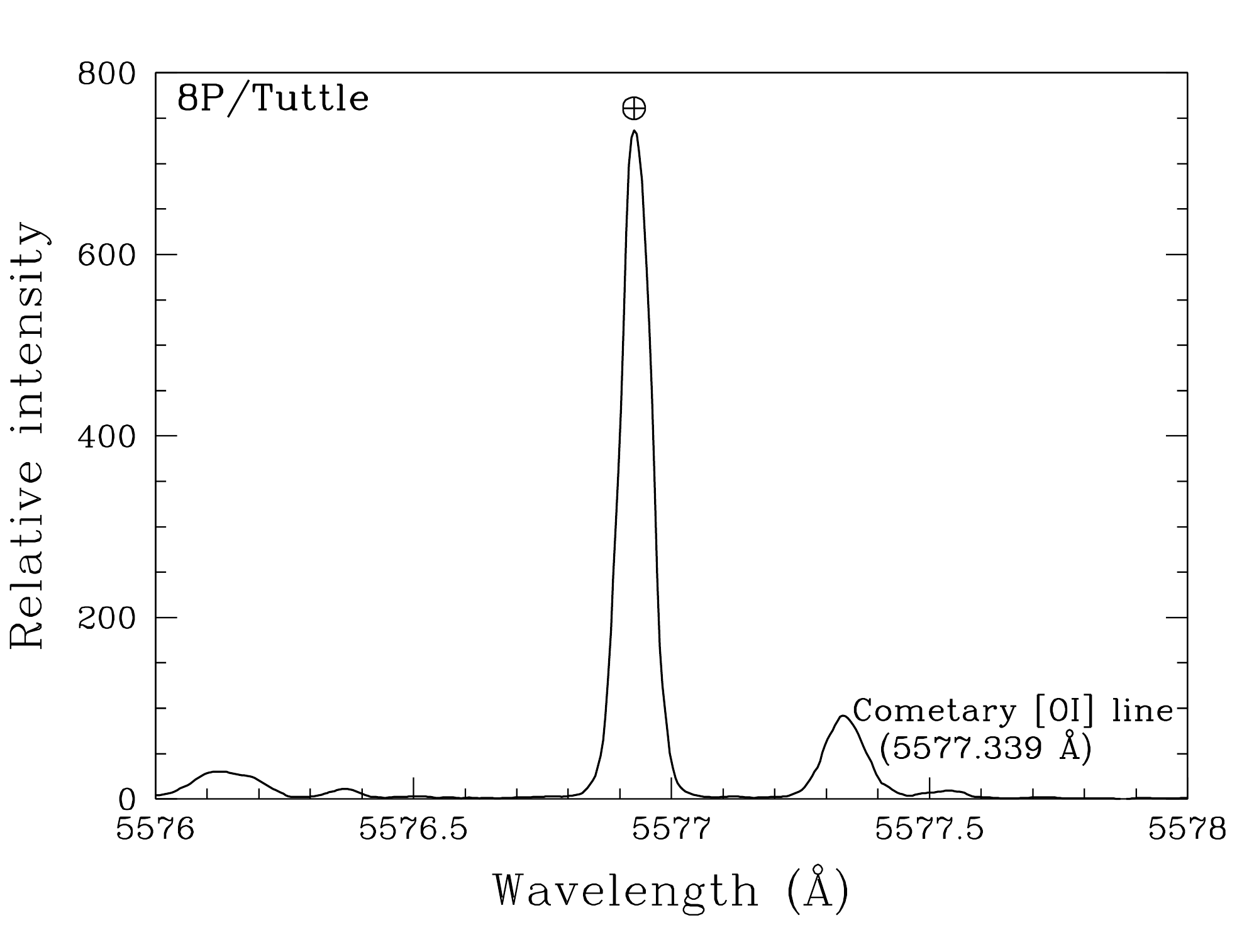}
\caption{The 6300.304 $\AA$, 6363.776 $\AA$ and 5577.339 $\AA$ oxygen lines (telluric ($\oplus$) and cometary) in a 3600s spectrum of comet 8P/Tuttle obtained in March 2008. Thanks to the high resolution of the UVES spectrograph, the cometary and telluric lines are well separated when the Doppler shift is larger than 15 km s$^{-1}$. These spectra are representative of the quality of the sample.}
\label{tuttler1}
\end{figure}

Our analysis is based on 12 comets listed in Table~\ref{comets}, considering the two components of 73P/Schwassmann-Wachmann 3 (B and C) as two comets. This sample is characterized by a large diversity. These comets have various dynamical origins (external, new, Jupiter family, Halley type) and are observed at different heliocentric distances (from 0.68 AU to 3.73 AU). Some are split comets like 73P or did a close approach to Earth. Others have been observed at large heliocentric distances like C/2001 Q4 and C/2009 P1. The observing material is made of a selection of 48 high signal-to-noise spectra obtained during the last ten years (\cite{Manfroid2009} ; \cite{Jehin2009} and references therein) with the cross-dispersed echelle spectrograph UVES \citep{Dekker2000}. A dichroic filter splits the light beam into two arms (a red one from 3000 to 5000 $\AA$ and a blue one from 4200 to 11000 $\AA$). In most observations, a slit of 0.45$^{\prime}$$^{\prime}$ $\times$ 11$^{\prime}$$^{\prime}$ was used. With such a slit, the resolving power is $R = \lambda/\Delta \lambda ~= ~110000$ in the red arm where the three [OI] lines are observed. All spectra were recorded with nearly the same instrumental setting. Such a high resolution is needed to isolate the [OI] emission lines from the telluric lines and other cometary lines (C$_{2}$, NH$_{2}$) (see Fig.\ref{tuttler1}). The slit was always centered on the comet nucleus. For a comet located at 1 AU from the Earth, the spatial area covered by the slit is approximatively 320 km by 8000 km. The size of the slit for each individual spectrum is given in Table~\ref{bigtable}. 
 
\begin{table*}
\centering
\tiny
\begin{supertabular}{p{4.7cm}p{1.8cm}p{1.2cm}p{1cm}p{1cm}p{1cm}p{1.1cm}p{0.9cm}p{1cm}}
\hline 
\hline
Comet & $Tp$ & $e$ & \emph{a} & $q$ & $i$ & $P$ & $T_{J}$ & Type~(L) \\
\hline
C/2002 V1 (NEAT) & 18-02-2003 & 0.99990 & 1010 & 0.10 & 82 & 32100 & 0.06 & \hfill{EXT} \\
C/2002 X5 (Kudo-Fujikawa) & 29-01-2003 & 0.99984 & 1175 & 0.19 & 94 & 40300 & -0.03 & \hfill{EXT} \\
C/2002 Y1 (Juels-Holvorcem) & 13-04-2003 & 0.99715 & 250.6 & 0.71 & 104 & 3967 & -0.23 & \hfill{EXT} \\
C/2001 Q4 (NEAT) & 16-05-2004 & 1.00069 & - & 0.96 & 100 & - & - & \hfill{NEW} \\
C/2002 T7 (LINEAR) & 23-04-2004 & 1.00048 & - & 0.61 & 161 & - & - & \hfill{NEW} \\
C/2003 K4 (LINEAR) & 14-10-2004 & 1.00030 & - & 1.02 & 134 & - & - & \hfill{NEW} \\
9P/Tempel 1 & 05-07-2005 & 0.51749 & 3.1 & 1.51 & 11 & 5.5 & 2.97 & \hfill{JF} \\
73P-C/Schwassmann-Wachmann 3 & 07-06-2006 & 0.69338 & 3.1 & 0.94 & 11 & 5.4 & 2.78 & \hfill{JF}  \\
73P-B/Schwassmann-Wachmann 3 & 08-06-2006 & 0.69350 & 3.1 & 0.94 & 11 & 5.4 & 2.78 & \hfill{JF}  \\
8P/Tuttle &  27-01-2008 & 0.81980 & 5.7 & 1.03 & 55 & 13.6 & 1.60 & \hfill{HF} \\
103P/Hartley 2 & 28-10-2010 & 0.69500 & 3.5 & 1.06 & 14 & 6.5 & 2.64 & \hfill{JF} \\
C/2009 P1 (Garradd) & 23-12-2011 & 1.00110 & - & 1.55 & 106 & - & - & \hfill{NEW}  \\
\hline
\end{supertabular}
\caption{\label{comets}Orbital characteristics of the 12 comets and their classification. $T_{p}$ is the epoch of the perihelion (dd-mm-yyyy), $e$ the eccentricity, $a$ is related to the semi-major axis of the orbit (AU), $q$ is the perihelion distance (AU), $i$ represents the inclination on the ecliptic (degrees), $P$ is the period (years), $T_{J}$ is the Tisserand parameter relative to Jupiter. The last data correspond to the classification of the comets according to \cite{Levison1996} : HF and JF mean "Halley Family" and "Jupiter family" and gather comets with short period (< 200 years) ; EXT and NEW are respectively for external comets ($a$<$10000$~AU) and new comets ($a$>$10000$~AU) which came from the Oort Cloud.}
\end{table*}

\begin{table*}[htbp]
\begin{center}
\tiny
\tablefirsthead{
\hline
\hline
Comet & JD - 2~450~000.5 & $r$~(AU) & $\dot{r}$~(km~s$^{-1}$) & $\Delta$~(AU) & $\dot{\Delta}$~(km~s$^{-1}$) & Exptime~(s) & Slit~($^{\prime \prime}\times^{\prime\prime}$) & \hfill{Slit~(km~$\times$~km)} \\
\hline
}
\tablelasttail{\hline}
\bottomcaption{\label{bigtable} Individual spectra. JD is denoted for Julian Day, $r$ is the heliocentric distance, $\Delta$ the geocentric distance. $\dot{r}$ and $\dot{\Delta}$ are respectively the heliocentric and geocentric velocities. Exptime corresponds to the exposure time in seconds. Slit in arc seconds gives the size of the entrance slit of the spectrograph. The last column provides in km the spatial area covered by the slit.}
\begin{supertabular}{p{3.4cm} >{\centering\arraybackslash}p{2cm} >{\centering\arraybackslash}p{0.5cm} >{\centering\arraybackslash}p{1cm} >{\centering\arraybackslash}p{0.7cm} >{\centering\arraybackslash}p{0.9cm} >{\centering\arraybackslash}p{1.2cm}>{\centering\arraybackslash}p{1.5cm} p{1.8cm}}
C/2002 V1 (NEAT) & 2 647.037 & 1.22 & -36.51 & 0.83 & 7.88  & 2100 &  0.45 $\times$ 11.00 & \hfill{271 $\times$ 6~622} \\
C/2002 V1 (NEAT) & 2 647.062 & 1.22 & -36.53 & 0.83 & 7.99 & 2100 &  0.45 $\times$ 11.00 & \hfill{271 $\times$ 6~622} \\
C/2002 V1 (NEAT) & 2 649.031 & 1.18 & -37.11 & 0.84 & 8.29 & 2100 &  0.45 $\times$ 11.00 & \hfill{274 $\times$ 6~702}   \\
C/2002 V1 (NEAT) & 2 649.056 & 1.18 & -37.11 & 0.84 & 8.33 & 1987 &  0.45 $\times$ 11.00 & \hfill{274 $\times$ 6~702}   \\
C/2002 V1 (NEAT) & 2 719.985 & 1.01 & 39.76 & 1.63 & 42.02 & 600 &  0.45 $\times$ 11.00 & \hfill{532 $\times$ 13~004}  \\
C/2002 X5 (Kudo-Fujikawa) & 2 705.017 & 1.06 & 37.01 & 0.99 & 29.34 & 1800 &  0.45 $\times$ 11.00 & \hfill{323 $\times$ 7~898}  \\
C/2002 X5 (Kudo-Fujikawa) & 2 705.039 & 1.07 & 37.00 & 0.99 & 29.40 & 1800 &  0.45 $\times$ 11.00 & \hfill{323 $\times$ 7~898}  \\
C/2002 X5 (Kudo-Fujikawa) & 2 705.060 & 1.07 & 36.99 & 0.99 & 29.45 & 1800 &  0.45 $\times$ 11.00 & \hfill{323 $\times$ 7~898}  \\
C/2002 Y1 (Juels-Holvorcem) & 2 788.395 & 1.14 & 24.09 & 1.56 & -7.24 & 1800 &  0.40 $\times$ 11.00 & \hfill{453 $\times$ 12~446}  \\
C/2002 Y1 (Juels-Holvorcem) & 2 788.416 & 1.14 & 24.09 & 1.56 & -7.21 & 1800 &  0.40 $\times$ 11.00 & \hfill{453 $\times$ 12~446}  \\
C/2002 Y1 (Juels-Holvorcem) & 2 789.394 & 1.16 & 24.18 & 1.55 & -7.21 & 1800 &  0.40 $\times$ 11.00 & \hfill{450 $\times$ 12~366} \\
C/2002 Y1 (Juels-Holvorcem) & 2 789.415 & 1.16 & 24.19 & 1.55 & -7.18 & 1800 &  0.40 $\times$ 11.00 & \hfill{450 $\times$ 12~366} \\
C/2001 Q4 (NEAT) & 2 883.293 & 3.73 & -18.80 & 3.45 & -25.42 & 4500 &  0.45 $\times$ 11.00 & \hfill{1~126 $\times$ 27~524}\\
C/2001 Q4 (NEAT) & 2 883.349 & 3.73 & -18.80 & 3.45 & -25.32 & 4500 &  0.45 $\times$ 11.00 & \hfill{1~126 $\times$ 27~524}\\
C/2001 Q4 (NEAT) & 2 889.236 & 3.67 & -18.91 & 3.36 & -23.67 & 7200 &  0.45 $\times$ 11.00 & \hfill{1~097 $\times$ 26~806} \\
C/2001 Q4 (NEAT) & 2 889.320 & 3.66 & -18.91 & 3.36 & -23.54 & 7200 &  0.45 $\times$ 11.00 & \hfill{1~097 $\times$ 26~806} \\
C/2002 T7 (LINEAR) & 3 131.421 & 0.68 & 15.83 & 0.61 & -65.62 & 1080 &  0.44 $\times$ 12.00 & \hfill{195 $\times$ 5~309} \\
C/2002 T7 (LINEAR) & 3 151.976 & 0.94 & 25.58 & 0.41 & 54.98 & 2678 &  0.30 $\times$ 12.00 & \hfill{89 $\times$ 3~568} \\
C/2002 T7 (LINEAR) & 3 152.036 & 0.94 & 25.59 & 0.42 & 55.20 & 1800 &  0.30 $\times$ 12.00 & \hfill{91 $\times$ 3~655} \\
C/2003 K4 (LINEAR) & 3 131.342 & 2.61 & -20.34 & 2.37 & -43.12 & 4946 &  0.80 $\times$ 11.00 & \hfill{1~375 $\times$ 18~908} \\
C/2003 K4 (LINEAR) & 3 132.343 & 2.59 & -20.35 & 2.35 & -42.95 & 4380 &  0.60 $\times$ 11.00 & \hfill{1~023 $\times$ 18~748} \\
C/2003 K4 (LINEAR) & 3 329.344 & 1.20 & 14.81 & 1.51 & -28.23 & 1500 &  0.44 $\times$ 12.00 & \hfill{482 $\times$ 13~142} \\
9P/Tempel 1 & 3 553.955 & 1.51 & -0.21 & 0.89 & 8.95 & 7200 &  0.44 $\times$ 12.00 & \hfill{284 $\times$ 7~746} \\
9P/Tempel 1 & 3 554.954 & 1.51 & -0.15 & 0.89 & 9.07 & 7200 &  0.44 $\times$ 12.00 & \hfill{284 $\times$ 7~746} \\
9P/Tempel 1 & 3 555.955 & 1.51 & -0.04 &  0.90 & 9.19 & 7200 &  0.44 $\times$ 12.00 & \hfill{287 $\times$ 7~833} \\
9P/Tempel 1 & 3 557.007 & 1.51 & 0.09 & 0.90 & 9.48 & 9600 &  0.44 $\times$ 12.00 & \hfill{287 $\times$ 7~833} \\
9P/Tempel 1 & 3 557.955 & 1.51 & 0.20 & 0.91 & 9.44 & 7500 &  0.44 $\times$ 12.00 & \hfill{290 $\times$ 7~920} \\
9P/Tempel 1 & 3 558.952 & 1.51 & 0.31 & 0.91 & 9.55 & 7500 &  0.44 $\times$ 12.00 & \hfill{290 $\times$ 7~920} \\
9P/Tempel 1 & 3 559.954 & 1.51 & 0.43 & 0.92 & 9.68 & 7500 &  0.44 $\times$ 12.00 & \hfill{293 $\times$ 8~007} \\
9P/Tempel 1 & 3 560.952 & 1.51 & 0.55 & 0.93 & 9.80 & 7800 &  0.44 $\times$ 12.00 & \hfill{297 $\times$ 8~094}\\
9P/Tempel 1 & 3 561.953 & 1.51 & 0.66 & 0.93 & 9.91 & 7200 &  0.44 $\times$ 12.00 & \hfill{297 $\times$ 8~094}\\
9P/Tempel 1 & 3 562.956 & 1.51 & 0.78 & 0.94 & 10.04 & 7200 & 0.44 $\times$ 12.00 &\hfill{300 $\times$ 8~181}\\
73P-C/SW 3 & 3 882.367 & 0.95 & -4.17 & 0.15 & 12.31 & 4800 &  0.60 $\times$ 12.00 & \hfill{65 $\times$ 1~305} \\
73P-B/SW 3 & 3 898.369 & 0.94 & 1.79 & 0.25 & 13.10 & 4800 &  0.60 $\times$ 12.00 & \hfill{109 $\times$ 2~176} \\
8P/Tuttle & 4 481.021 & 1.04 & -4.29 & 0.36 & 21.64 & 3600 &  0.44 $\times$ 10.00 & \hfill{115 $\times$ 2~611} \\
8P/Tuttle & 4 493.018 & 1.03 & 0.40 & 0.52 & 24.72 & 3900 &  0.44 $\times$ 10.00 & \hfill{166 $\times$ 3~771} \\
8P/Tuttle & 4 500.017 & 1.03 & 3.16 & 0.62 & 24.16 & 3900 &  0.44 $\times$ 10.00 & \hfill{198 $\times$ 4~497} \\
103P/Hartley 2 & 5 505.288 & 1.06 & 2.53 & 0.16 & 7.08 & 2900 &  0.44 $\times$ 12.00 & \hfill{51 $\times$ 1~393} \\
103P/Hartley 2 & 5 505.328 & 1.06 & 2.55 & 0.16 & 7.19 & 3200 &  0.44 $\times$ 12.00 & \hfill{51 $\times$ 1~393} \\
103P/Hartley 2 & 5 510.287 & 1.07 & 4.05 & 0.18 & 7.96 & 2900 &  0.44 $\times$ 12.00 & \hfill{57 $\times$ 1~567} \\
103P/Hartley 2 & 5 510.328 & 1.07 & 4.07 & 0.18 & 8.07 & 3200 &  0.44 $\times$ 12.00 & \hfill{57 $\times$ 1~567} \\
103P/Hartley 2 & 5 511.363 & 1.08 & 4.37 & 0.19 & 8.27 & 900 &  0.44 $\times$ 12.00 & \hfill{60 $\times$ 1~654} \\
C/2009 P1 (Garradd) & 5 692.383 & 3.25 & -16.91 & 3.50 & -44.66 & 3600 &  0.44 $\times$ 12.00 & \hfill{1~117 $\times$ 30~461}\\
C/2009 P1 (Garradd) & 5 727.322 & 2.90 & -16.89 & 2.57 & -46.38 & 3600 &  0.44 $\times$ 12.00 & \hfill{820 $\times$ 22~367} \\
C/2009 P1 (Garradd) & 5 767.278 & 2.52 & -16.46 & 1.64 & -29.26 & 1800 &  0.44 $\times$ 12.00 & \hfill{523 $\times$ 14~273}  \\ 
C/2009 P1 (Garradd) & 5 813.991 & 2.09 & -14.82 & 1.47 & 14.79 & 4800 &  0.44 $\times$ 12.00 & \hfill{469 $\times$ 12~794} \\
C/2009 P1 (Garradd) & 5 814.974 & 2.08 & -14.77 & 1.48 & 15.31 & 4800 &  0.44 $\times$ 12.00 & \hfill{472 $\times$ 12~881} \\
C/2009 P1 (Garradd) & 5 815.982 & 2.07 & -14.71 & 1.49 & 15.84 & 4800 &  0.44 $\times$ 12.00 & \hfill{475 $\times$ 12~968} \\
\hline
\end{supertabular}
\end{center}
\end{table*}
\normalsize

\section{Data reduction}

\begin{figure}[h]
\centerline{\includegraphics[width=\columnwidth]{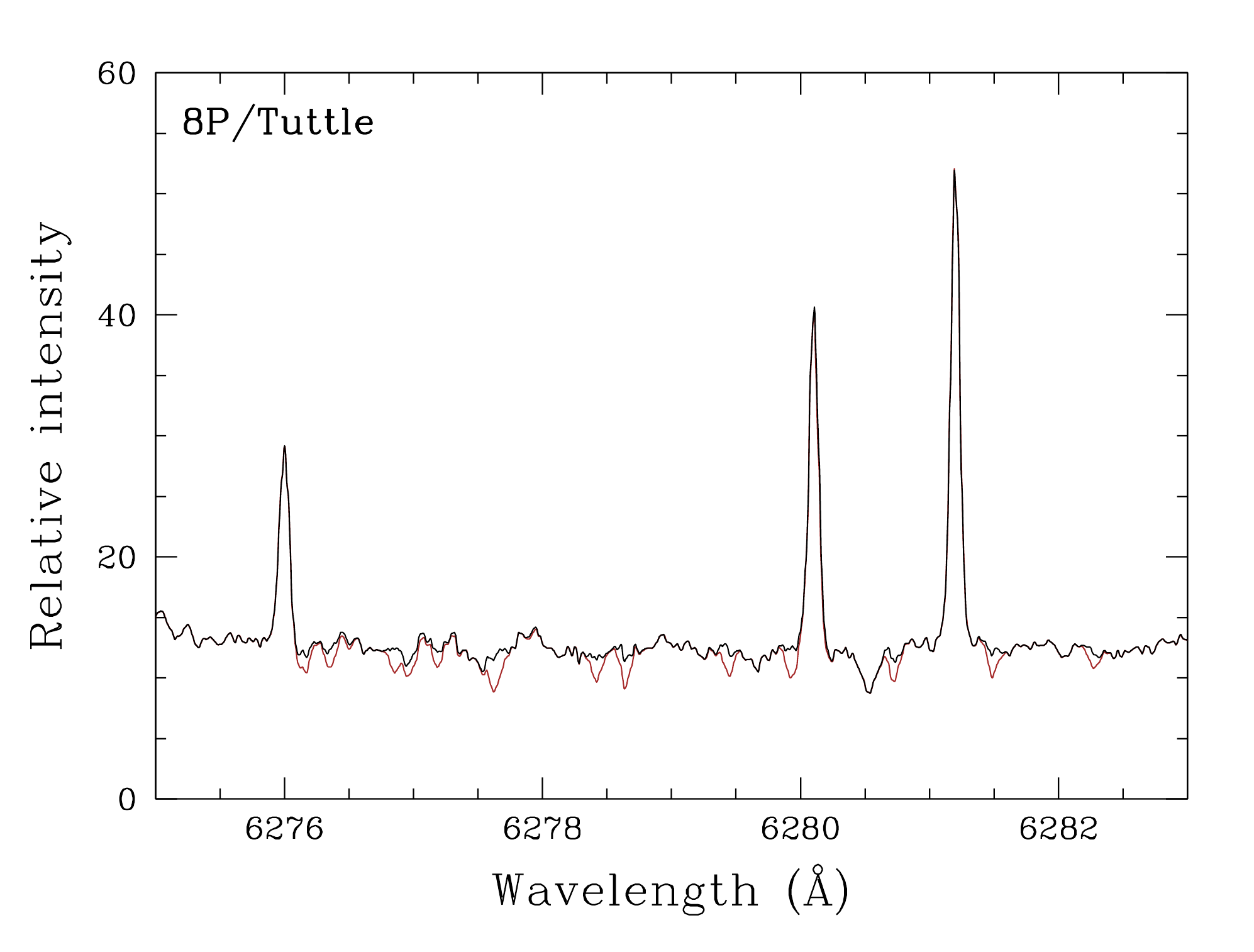}}
\caption{The telluric absorption line correction in the vicinity of the red [OI] lines. The black spectrum has been corrected (thick line), the red one is uncorrected.}
\label{telluric}
\end{figure}

The 2D spectra were reduced with the UVES pipeline reduction program.
For data obtained until 2008, the UVES pipeline (version 2.8.0) was used within the ESO-MIDAS environment to extract, for each order separately, the 2D spectra, bias-subtracted, flat-fielded and wavelength calibrated. ESO-MIDAS is a special software containing packages to reduce ESO data\footnote{http://www.eso.org/sci/software/esomidas/}. For a given setting, the orders were then merged using a weighting scheme to correct for the blaze function, computed from high S/N master flat-fields reduced in exactly the same way. This procedure leads to a good order merging. The post-2008 observations were reduced with the gasgano CPL\footnote{http://www.eso.org/sci/software/gasgano/} interface (version 2.4.3) of the re-furbished UVES pipeline that directly provides accurately merged 2D spectra. 1D spectra were extracted by averaging the 2D ones, with cosmic rejection.  Standard stars were similarly reduced and combined to derive instrumental response functions used to correct the 1D spectra. The wavelength calibration was computed using Thorium Argon (Th-Ar) spectra. The calibration revealed small position shifts of the lines due to the fact that the Th-Ar spectra were usually taken in the day and not right after the comet observations. This shift was removed using 9 cometary lines of NH$_{2}$ in the vicinity of the red [OI] lines and for which laboratory wavelengths are well known. \newline
Before doing any measurement, the telluric absorption lines were removed and the solar continuum contribution subtracted using the BASS\footnote{http://bass2000.obspm.fr/solar$\_$spect.php} and Kurucz\footnote{http://kurucz.havard.edu/sun/irradiance2005/irradthu.dat} spectra. These two spectra are solar spectra, with atmospheric absorption lines for the BASS spectrum and without for the Kurucz one (2005). Around the red doublet line at 6300~\AA, there are indeed telluric absorption lines mostly due to O$_{2}$ molecules in the Earth's atmosphere which could lead to an underestimate of the forbidden oxygen line intensity (see Fig.~\ref{telluric}).

\begin{figure}[h]
\centerline{\includegraphics[width=\columnwidth]{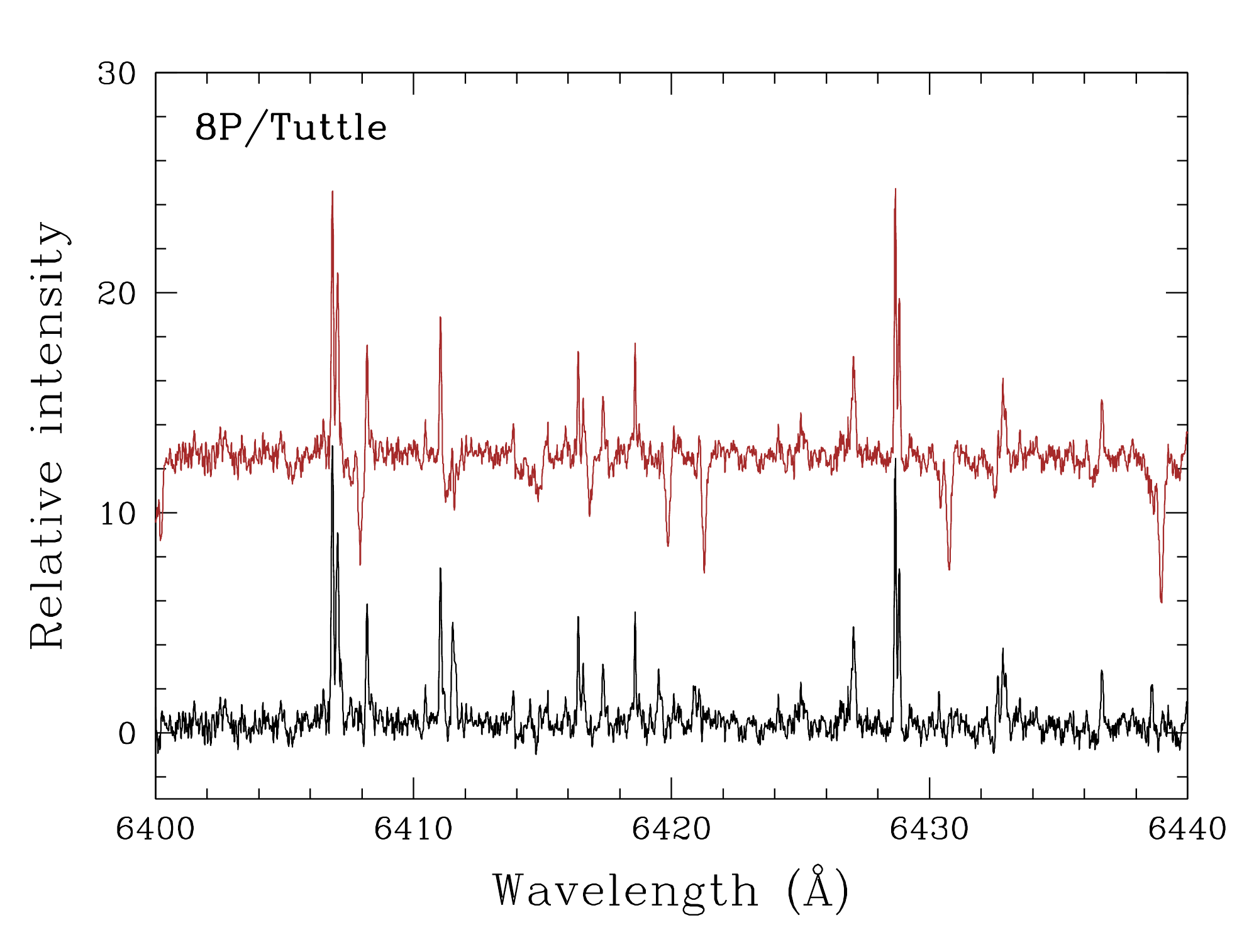}}
\caption{The solar continuum has been removed in the lower spectrum (black) while it is present in the upper one (in red). It corresponds to the final spectrum on which the measurements of the [OI] are made.}
\label{solar}
\end{figure}

The comet Solar continuum has three contributions. First and foremost is the one due to the reflection of sunlight on the dust particles in the coma. Another one appears when the observations are made in or close to the twilight. A third contribution is, in a few cases, the background radiation by the Moon. 
 To remove these contaminations, we Doppler shifted the BASS spectrum to the proper values and scaled the intensities until the solar features are completely removed. This provides the final spectrum such as the one presented in Fig.~\ref{solar} for comet 8P/Tuttle, free of any solar and telluric lines.
 
\section{Data analysis}

After the reduction and the correction of the data, we measured both the intensity and the FWHM of the three forbidden oxygen lines by fitting a gaussian profile using the IRAF\footnote{IRAF is a tool for the reduction and the analysis of astronomical data (http://iraf.noao.edu).} software. 
The observed width is the convolution of the instrumental profile with the natural width :

\tiny
\begin{eqnarray}
\label{fwhm}
\rm{FWHM_{observed} (\lambda)} = \sqrt{\rm{FWHM^{2}_{intrinsic} (\lambda)} + \rm{FWHM^{2}_{instrumental}(\lambda)}}~,
\end{eqnarray}
\normalsize
where the instrumental width corresponds to the width of the Th-Ar lines.

\paragraph{}

These measurements can only be performed with best accuracy when the cometary oxygen lines are well separated from the telluric [OI] lines, i.e, when the geocentric velocitiy of the comet exceeds 15 km~s$^{-1}$. Using the deblend function of the {\it splot} package in IRAF, we could measure the [OI] lines for Doppler shifts as low as 7~km~s$^{-1}$.

\section{Results}

We have measured the two ratios and the FWHM of the three [OI]  lines. To reach such goals, a relative calibration using the well known UVES instrumental response curve provided by ESO was enough. Nevertheless, we will investigate how to produce absolute flux calibrated spectra in order to derive production rates of the oxygen parent species in a future work. Therefore, in this current study, the production rates of H$_{2}$O to assess the activity of the comet have been taken from literature.

\subsection{Intensity ratios}

\begin{table*}[htbp]
\begin{center}
\tiny
\tablefirsthead{
\hline
\hline
Comet & JD - 2~450~000.5 & r~(AU) & \multicolumn{3}{c}{Intensity (ADU)} & $\frac{I_{6300}}{I_{6364}}$ & $\frac{I_{5577}}{I_{6300} + I_{6364}}$ \\
 & & & 5577.339~\r{A} & 6300.304~\r{A} & 6363.776~\r{A} & & \\
\hline
}
\tablelasttail{\hline}
\bottomcaption{\label{resultat} Intensities given in ADU (Analog-to-Digital Units) and measured line ratios for the three forbidden oxygen lines in the spectra of each comet. For the third spectrum of comets C/2001 Q4 and 9P, the green line could not be measured because of a contamination by a cosmic ray.}
\begin{supertabular}{p{4cm} >{\centering\arraybackslash}p{2cm} >{\centering\arraybackslash}p{1.1cm} >{\centering\arraybackslash}p{1.1cm} >{\centering\arraybackslash}p{1.1cm} >{\centering\arraybackslash}p{1.2cm} >{\centering\arraybackslash}p{1.2cm} p{1cm}}
C/2002 V1 (NEAT) & 2 647.037 & 1.22 & 2348 & 27539 & 8875 & 3.10 & \hfill{0.065} \\
C/2002 V1 (NEAT) & 2 647.062 & 1.22 & 2255 & 26790 & 8875 & 3.02 & \hfill{0.063} \\
C/2002 V1 (NEAT) & 2 649.031 & 1.18 & 3220 & 35672 & 11704 & 3.05 & \hfill{0.068} \\
C/2002 V1 (NEAT) & 2 649.056 & 1.18 & 3137 & 33738 & 11105 & 3.04 & \hfill{0.070} \\
C/2002 V1 (NEAT) & 2 719.985 & 1.01 & 8657 & 68162 & 22381 & 3.05 & \hfill{0.096} \\
C/2002 X5 (Kudo-Fujikawa) & 2 705.017 & 1.06 & 4016 & 31491 & 10589 & 2.97 & \hfill{0.095} \\
C/2002 X5 (Kudo-Fujikawa) & 2 705.039 & 1.07 & 4430 & 32822 & 10997 & 2.98 & \hfill{0.101} \\
C/2002 X5 (Kudo-Fujikawa) & 2 705.060 & 1.07 & 4249 & 32510 & 10816 & 3.01 & \hfill{0.098} \\
C/2002 Y1 (Juels-Holvorcem) & 2 788.395 & 1.14 & 6223 & 50960 & 17387 & 2.93 & \hfill{0.091} \\
C/2002 Y1 (Juels-Holvorcem) & 2 788.416 & 1.14 & 6386 & 54683 & 18167 & 3.01 & \hfill{0.088} \\
C/2002 Y1 (Juels-Holvorcem) & 2 789.394 & 1.16 & 3942 & 25792 & 8659 & 2.98 & \hfill{0.114} \\
C/2002 Y1 (Juels-Holvorcem) & 2 789.415 & 1.16 & 4254 & 31262 & 10398 & 3.01 & \hfill{0.102} \\
C/2001 Q4 (NEAT) & 2 883.293 & 3.73 & 143 & 324 & 110 & 2.95 & \hfill{0.329} \\
C/2001 Q4 (NEAT) & 2 883.349 & 3.73 & 151 & 353 & 110 & 3.22 & \hfill{0.326} \\
C/2001 Q4 (NEAT) & 2 889.236 & 3.67 & - & 354 & 112 & 3.17 & \hfill{-} \\
C/2001 Q4 (NEAT) & 2 889.320 & 3.66 & 261 & 677 & 193 & 3.51 & \hfill{0.300} \\
C/2002 T7 (LINEAR) & 3 131.421 & 0.68 & 116522 & 642304 & 211120 & 3.04 & \hfill{0.137} \\
C/2002 T7 (LINEAR) & 3 151.976 & 0.94 & 50544 & 365664 & 104187 & 3.51 & \hfill{0.108}  \\
C/2002 T7 (LINEAR) & 3 152.036 & 0.94 & 40165 & 327184 & 96554 & 3.39 & \hfill{0.095} \\
C/2003 K4 (LINEAR) & 3 131.342 & 2.61 & 464 & 3609 & 1172 & 3.08 & \hfill{0.097}  \\
C/2003 K4 (LINEAR) & 3 132.343 & 2.59 & 561 & 5027 & 1607 & 3.13 & \hfill{0.085} \\
C/2003 K4 (LINEAR) & 3 329.344 & 1.20 & 9244 & 94723 & 29640 & 3.20 & \hfill{0.074} \\
9P/Tempel 1 & 3 553.955 & 1.51 & 230 & 3178 & 1039 & 3.06 & \hfill{0.054} \\
9P/Tempel 1 & 3 554.954 & 1.51 & 172 & 3153 & 1051 & 3.00 & \hfill{0.041} \\
9P/Tempel 1 & 3 555.955 & 1.51 & - & 3894 & 1297 & 3.00 & \hfill{-} \\
9P/Tempel 1 & 3 557.007 & 1.51 & 341 & 5618 & 1842 & 3.05 & \hfill{0.046} \\
9P/Tempel 1 & 3 557.955 & 1.51 & 223 & 5154 & 1595 & 3.23 & \hfill{0.033} \\
9P/Tempel 1 & 3 558.952 & 1.51 & 277 & 5267 & 1629 & 3.23 & \hfill{0.040} \\
9P/Tempel 1 & 3 559.954 & 1.51 & 291 & 4618 & 1440 & 3.21 & \hfill{0.048} \\
9P/Tempel 1 & 3 560.952 & 1.51 & 388 & 4984 & 1571 & 3.17 & \hfill{0.059} \\
9P/Tempel 1 & 3 561.953 & 1.51 & 211 & 4281 & 1347 & 3.18 & \hfill{0.037} \\
9P/Tempel 1 & 3 562.956 & 1.51 & 142 & 4162 & 1281 & 3.25 & \hfill{0.026} \\
73P-C/Schwassmann-Wachmann 3 & 3 882.367 & 0.95 & 16505 & 115170 & 35526 & 3.24 & \hfill{0.110} \\
73P-B/Schwassmann-Wachmann 3 & 3 898.369 & 0.94 & 4890 & 33987 & 11226 & 3.03 & \hfill{0.108} \\
8P/Tuttle & 4 481.021 & 1.04 & 1779 & 28517 & 9385 & 3.04 & \hfill{0.047} \\
8P/Tuttle & 4 493.018 & 1.03 & 4437 & 74360 & 24565 & 3.03 & \hfill{0.045} \\
8P/Tuttle & 4 500.017 & 1.03 & 4179 & 64979 & 21424 & 3.03 & \hfill{0.048}  \\
103P/Hartley 2 & 5 505.288 & 1.06 & 45 & 373 & 120 & 3.12 & \hfill{0.092} \\
103P/Hartley 2 & 5 505.328 & 1.06 & 50 & 456 & 144 & 3.22 & \hfill{0.100} \\
103P/Hartley 2 & 5 510.287 & 1.07 & 51 & 390 & 121 & 3.24 & \hfill{0.073} \\
103P/Hartley 2 & 5 510.328 & 1.07 & 41 & 425 & 131 & 3.17 & \hfill{0.083} \\
103P/Hartley 2 & 5 511.363 & 1.08 & 48 & 395 & 125 & 3.17 & \hfill{0.093} \\
C/2009 P1 (Garradd) & 5 692.383 & 3.25 & 180 & 652 & 224 & 2.92 & \hfill{0.205} \\
C/2009 P1 (Garradd) & 5 727.322 & 2.90 & 129 & 693 & 205 & 3.38 & \hfill{0.143} \\
C/2009 P1 (Garradd) & 5 767.278 & 2.52 & 1110 & 8276 & 2652 & 3.12 & \hfill{0.102} \\ 
C/2009 P1 (Garradd) & 5 813.991 & 2.09 & 1594 & 17035 & 5666 & 3.01 & \hfill{0.070} \\
C/2009 P1 (Garradd) & 5 814.974 & 2.08 & 1571 & 16873 & 5489 & 3.07 & \hfill{0.070} \\
C/2009 P1 (Garradd) & 5 815.982 & 2.07 & 1506 & 17557 & 5780 & 3.04 & \hfill{0.065} \\
\end{supertabular}
\end{center}
\end{table*}
\normalsize

\begin{table}
\centering
\begin{tabular}{l l l l}
\hline
\hline
Comet & N & $\frac{I_{6300}}{I_{6364}}$ &  $\frac{I_{5577}}{I_{6300} + I_{6364}}$  \\
\hline
C/2002 V1 (NEAT) & 5 & 3.05 $\pm$ 0.03 & 0.07 $\pm$ 0.01 \\
C/2002 X5 (Kudo-Fujikawa) & 3 & 2.99 $\pm$ 0.02 & 0.10 $\pm$ 0.003 \\
C/2002 Y1 (Juels-Holvorcem) & 4 & 2.98 $\pm$ 0.04 & 0.10 $\pm$ 0.01 \\
C/2001 Q4 (NEAT) & 4 & 3.21 $\pm$ 0.23 & 0.32 $\pm$ 0.02 \\
C/2002 T7 (LINEAR) & 3 & 3.31 $\pm$ 0.18 & 0.15 $\pm$ 0.05 \\
C/2003 K4 (LINEAR) & 3 & 3.14 $\pm$ 0.06 & 0.09 $\pm$ 0.01 \\
9P/Tempel 1 & 10 & 3.14 $\pm$ 0.10 & 0.04 $\pm$ 0.01 \\
73P-C/SW 3 & 1 & 3.25  & 0.11 \\
73P-B/SW 3 & 1 & 3.03 & 0.11\\
8P/Tuttle & 3 & 3.03 $\pm$ 0.01 & 0.05 $\pm$ 0.01 \\
103P/Hartley 2 & 5 & 3.15 $\pm$ 0.07 & 0.08 $\pm$ 0.02 \\
C/2009 P1 (Garradd) & 2 & 3.15 $\pm$ 0.33 & 0.017 $\pm$ 0.04 \\
C/2009 P1 (Garradd) & 3 & 3.04 $\pm$ 0.03 & 0.07 $\pm$ 0.003 \\
\hline
\end{tabular}
\caption{\label{moyenne} Average values of the line ratios for each comet. The errors listed are  the standard deviation of the N spectra available for each comet. 73P-C/SW3 and 73P-C/SW3 have no error because we obtained only one spectrum for these comets.}
\end{table}

\begin{figure}[h]
\centerline{\includegraphics[width=\columnwidth]{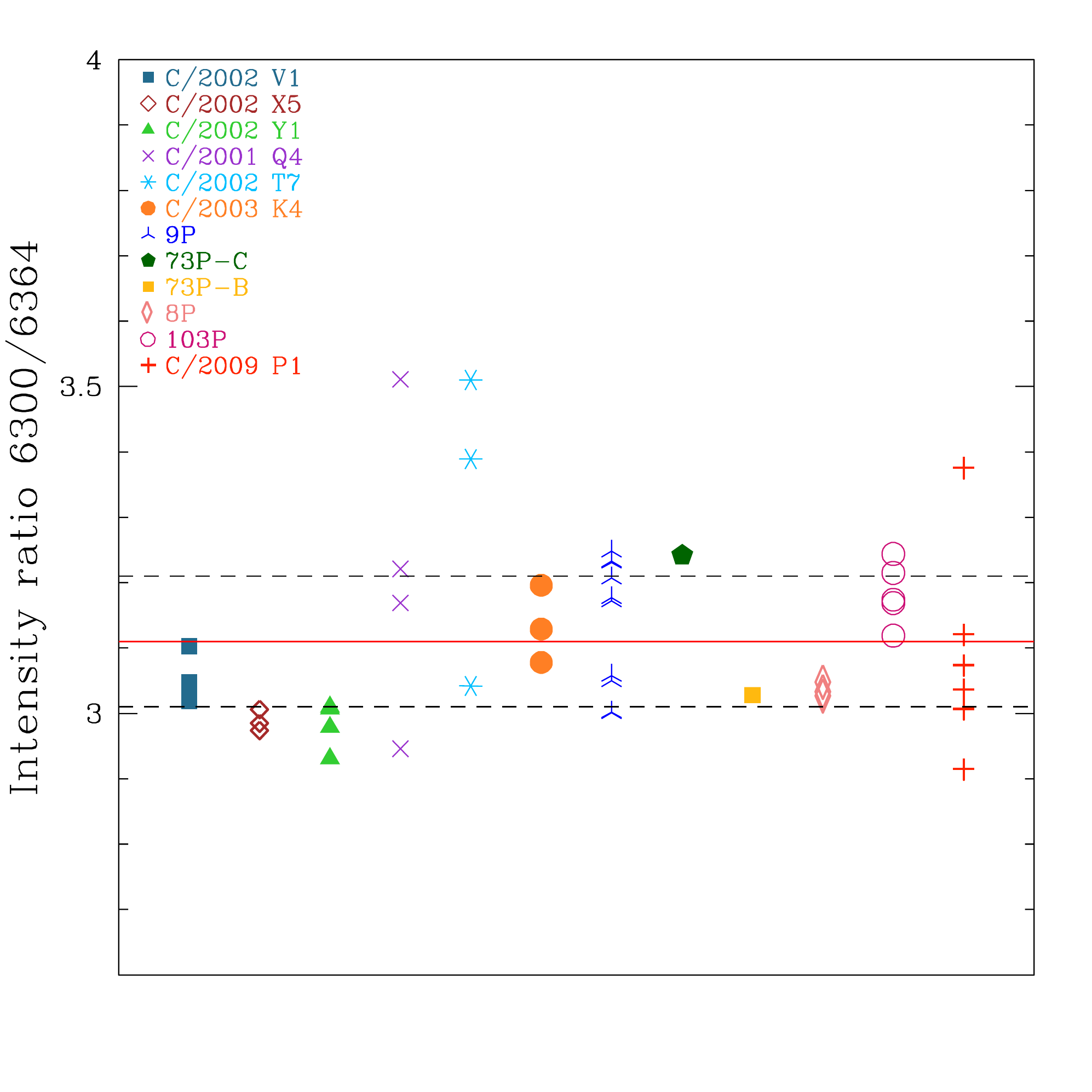}}
\caption{The red doublet ratio for each comet. The spectra of a given comet have the same symbols. The average value of the sample (3.11~$\pm$~0.10) is shown by the solid line. The dashed lines show the standard deviation ($\sigma$).}
\label{redratio}
\end{figure}

\begin{figure}[h]
\centerline{\includegraphics[width=\columnwidth]{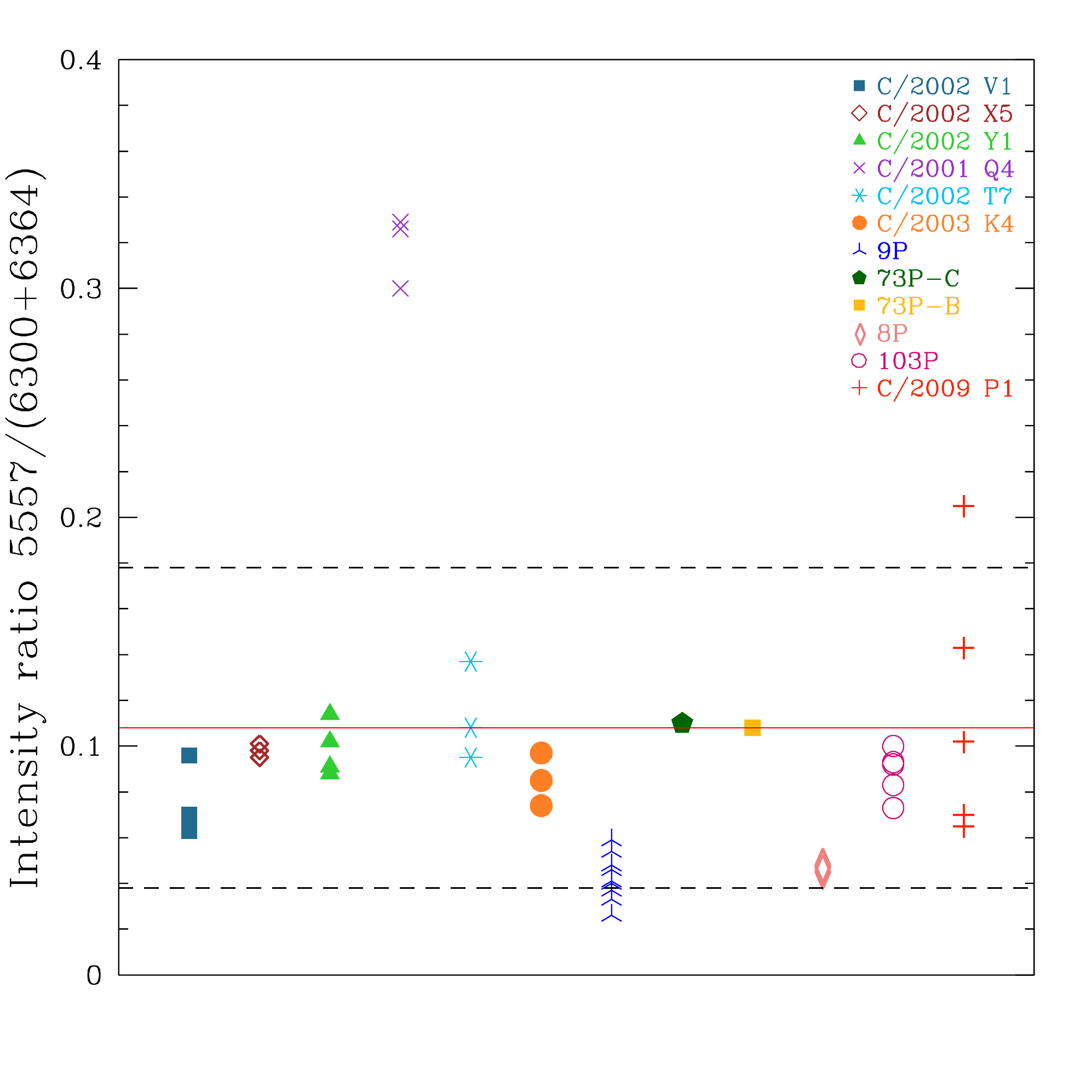}}
\caption{Same figure as Fig.~\ref{redratio} for the G/R ratio. The mean value is 0.11~$\pm$~0.07. Note the data points for comet Q4 (NEAT) at 3.7 AU.}
\label{greenratio}
\end{figure}

The two measured intensity ratios, $I_{6300}/I_{6364}$ and G/R~=~$I_{5577}/(I_{6300}+I_{6364})$, are displayed for each spectrum in Fig.~\ref{redratio} and Fig.~\ref{greenratio} respectively. When the collisional quenching is neglected, the intensity of a line is given by \citep{Festou1981} :
\begin{eqnarray}
\label{intensity}
I = ~~ \tau_{p}^{-1} ~~ \alpha ~~ \beta ~~ N ~~\rm{photons}~\rm{cm}^{-2}~\rm{s}^{-1}
\end{eqnarray}
where $\tau_{p}$ is the photo-dissociative lifetime of the parent, $\alpha$ is the yield of photo-dissociation, $\beta$ corresponds to the branching ratio for the transition and $N$ is the column density of the parent. Table~\ref{resultat} lists the line intensities $I$ (in ADU) and the ratios for all the spectra. Table~\ref{moyenne} shows the standard deviation ($\sigma$) of the ratio for all the spectra of a given comet. Since usually most of the spectra of a same comet are obtained during a short interval of time at similar heliocentric and geocentric distances as well as the same exposure time, the standard deviation of these measurements provides a good estimate of the error. Indeed this error includes the photon noise and errors coming from the telluric absorption and solar continuum corrections. \\ 
The red doublet ratio is remarkably similar for all comets. The average value over the whole sample of comets is  $3.11~\pm~0.10$. The error corresponds to the standard deviation ($\sigma$). The ratio is in agreement with the value of $3.09~\pm~0.12$ obtained by \cite{Cochran2008} from a sample of 8 comets but with a better accuracy. Since both red lines are transitions from the same level to the ground state, $\tau_{p}$, $\alpha$ and $N$ are equal in Eq.~(\ref{intensity}) and the intensity ratio $I_{6300}/I_{6364}$ is equivalent to the branching ratio $\beta_{6300}/\beta_{6364}$. Our results are indeed in very good agreement with the theoretical value of the branching ratio of $3.096$ computed for the quantum mechanics \citep{Galavis1997}. \cite{Storey2000} published a new theoretical value of $2.997$ taking into account relativistic effects which should be an improvement over previous determinations. Higher value derived from cometary spectra could point a systematic error due to a small blend of the 6300.304 \AA~[OI] line with an unidentified cometary feature. In order to check if such systematics is present, we first took the average red line ratio of the three comets at large heliocentric distances (C/2001 Q4, C/2003 K4 and C/2009 P1) since most of the fluorescence lines and then their contamination disappear far from the Sun. We obtained an average ratio of 3.13 $\pm$ 0.07. Secondly, we computed the the same ratio but this time for the terrestrial nightglow and found a ratio equal to 3.20~$\pm$~0.05 obtained again at the largest cometary distances. 
These measurements remain in better agreement with the \cite{Galavis1997} value but the error are too large to discard \cite{Storey2000} and we cannot exclude a systematic error.
\paragraph{}
The G/R ratio has an average value of $0.11~\pm~0.07$  for the complete sample (see Fig.~\ref{greenratio}).This result is in agreement with the value of $0.09~\pm~0.01$ found by \cite{Cochran2008}. The dispersion is higher because it includes comets at large heliocentric distances which have different G/R ratio. If we take only comets at < $2.5$~AU into account, the average and dispersion are equal to $0.09 \pm 0.02$. This leads to the conclusion that H$_{2}$O is the main parent molecule producing oxygen atoms according to \cite{Bhardwaj2012} values (Table~\ref{bhardwaj}).

\paragraph{}

\begin{figure}[h]
\centerline{\includegraphics[width=\columnwidth]{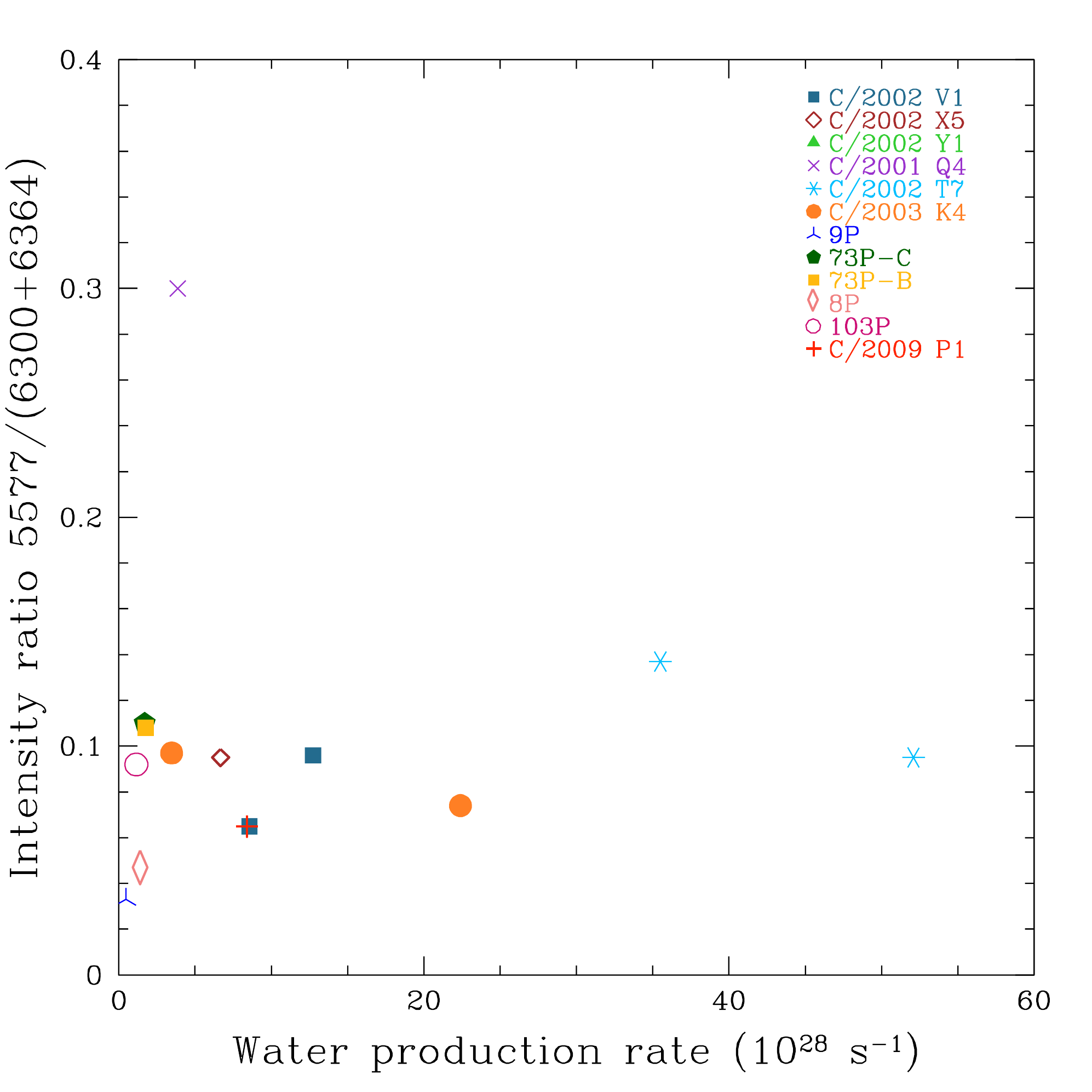}}
\caption{The G/R ratio against the water production rate. No trend appears. This analysis shows that the quenching can be neglected for our study.}
\label{greenratio_wp}
\end{figure}

In order to see if the quenching could play a role on our results, we computed the G/R ratio with respect to the water production rates (Fig.~\ref{greenratio_wp}). The water production rates of each comet has been compiled from the literature (cf. Table~\ref{waterp}). As those comets have been observed in different conditions (of production rates and heliocentric distances), the lack of trend in this analysis suggests that quenching is negligible and does not significantly affect our main conclusions. However, we have started to work on a model based on the \cite{Bhardwaj2012} paper and on Monte Carlo simulations to estimate the quenching. A careful computation of the quenching is not an easy task and we will give our new results, including a detailed discussion of the quenching, in a forthcoming paper.

\subsection{Line widths}

\begin{table*}[htbp]
\begin{center}
\tiny
\tablefirsthead{
\hline
\hline
Comet & JD - 2~450~000.5 & r~(AU) & \multicolumn{3}{c}{FWHM$_{\rm{observed}}$~($\AA$)} & \multicolumn{3}{c}{FWHM$_{\rm{intrinsic}}$~(km~s$^{-1}$)}  \\
 & & & 5577.339~\r{A} & 6300.304~\r{A} & 6363.776~\r{A} & 5577.339~\r{A} & 6300.304~\r{A} &  \hfill{6363.776~\r{A}} \\
\hline
}
\tablelasttail{\hline}
\bottomcaption{\label{resultat2} Measured observed FWHM ($\AA$) and intrinsic velocity widths (km s$^{-1}$) for the three forbidden oxygen lines in the spectra of each comet. For the third spectrum of comets C/2001 Q4 and 9P, the green line could not be used because of the contamination by a strong cosmic ray.}
\begin{supertabular}{p{3.95cm} >{\centering\arraybackslash}p{2cm} >{\centering\arraybackslash}p{0.8cm}>{\centering\arraybackslash}p{1.2cm} >{\centering\arraybackslash}p{1.2cm} >{\centering\arraybackslash}p{1.2cm} >{\centering\arraybackslash}p{1.2cm} >{\centering\arraybackslash}p{1.2cm} p{1.5cm}}
C/2002 V1 (NEAT) & 2 647.037 & 1.22 & 0.092 & 0.098 & 0.094 & 2.08 & 1.57 & \hfill{1.40} \\
C/2002 V1 (NEAT) & 2 647.062 & 1.22 & 0.089 & 0.095 & 0.095 & 1.95 & 1.42 & \hfill{1.44} \\
C/2002 V1 (NEAT) & 2 649.031 & 1.18 & 0.096 & 0.096 & 0.095 & 2.23 & 1.45 & \hfill{1.42} \\
C/2002 V1 (NEAT) & 2 649.056 & 1.18 & 0.093 & 0.095 & 0.094 & 2.11 & 1.40 & \hfill{1.38} \\
C/2002 V1 (NEAT) & 2 719.985 & 1.01 & 0.108 & 0.098 & 0.096 & 2.76 & 1.54 & \hfill{1.49} \\
C/2002 X5 (Kudo-Fujikawa) & 2 705.017 & 1.06 & 0.104 & 0.094 & 0.093 & 2.64 & 1.46 & \hfill{1.50} \\
C/2002 X5 (Kudo-Fujikawa) & 2 705.039 & 1.07 & 0.104 & 0.095 & 0.095 & 2.61 & 1.54 & \hfill{1.60} \\
C/2002 X5 (Kudo-Fujikawa) & 2 705.060 & 1.07 & 0.104 & 0.095 & 0.094 & 2.64 & 1.50 & \hfill{1.54} \\
C/2002 Y1 (Juels-Holvorcem) & 2 788.395 & 1.14 & 0.106 & 0.092 & 0.091 & 2.88 & 1.52 & \hfill{1.47} \\
C/2002 Y1 (Juels-Holvorcem) & 2 788.416 & 1.14 & 0.103 & 0.092 & 0.091 & 2.75 & 1.54 & \hfill{1.49} \\
C/2002 Y1 (Juels-Holvorcem) & 2 789.394 & 1.16 & 0.113 & 0.095 & 0.094 & 3.09 & 1.59 & \hfill{1.57} \\
C/2002 Y1 (Juels-Holvorcem) & 2 789.415 & 1.16 & 0.112 & 0.095 & 0.094 & 3.03 & 1.61 & \hfill{1.59} \\
C/2001 Q4 (NEAT) & 2 883.293 & 3.73 & 0.101 & 0.116 & 0.117 & 2.51 & 2.39 & \hfill{2.39} \\
C/2001 Q4 (NEAT) & 2 883.349 & 3.73 & 0.097 & 0.123 & 0.125 & 2.31 & 2.63 & \hfill{2.69} \\
C/2001 Q4 (NEAT) & 2 889.236 & 3.67 & - & 0.119 & 0.121 & - & 2.49 & \hfill{2.52} \\
C/2001 Q4 (NEAT) & 2 889.320 & 3.66 & 0.102 & 0.125 & 0.117 & 2.55 & 2.75 & \hfill{2.36} \\
C/2002 T7 (LINEAR) & 3 131.421 & 0.68 & 0.115 & 0.099 & 0.100 & 3.12 & 1.82 & \hfill{1.80} \\
C/2002 T7 (LINEAR) & 3 151.976 & 0.94 & 0.101 & 0.089 & 0.090 & 2.84 & 1.70 & \hfill{1.71}  \\
C/2002 T7 (LINEAR) & 3 152.036 & 0.94 & 0.100 & 0.088 & 0.090 & 2.81 & 1.66 & \hfill{1.72} \\
C/2003 K4 (LINEAR) & 3 131.342 & 2.61 & 0.131 & 0.135 & 0.136 & 2.51 & 1.81 & \hfill{1.66}  \\
C/2003 K4 (LINEAR) & 3 132.343 & 2.59 & 0.106 & 0.108 & 0.105 & 2.76 & 2.12 & \hfill{1.94} \\
C/2003 K4 (LINEAR) & 3 329.344 & 1.20 & 0.097 & 0.094 & 0.096 & 2.38 & 1.52 & \hfill{1.52} \\
9P/Tempel 1 & 3 553.955 & 1.51 & 0.110 & 0.096 & 0.093 & 2.91 & 1.61 & \hfill{1.43} \\
9P/Tempel 1 & 3 554.954 & 1.51 & 0.091 & 0.095 & 0.094 & 2.14 & 1.59 & \hfill{1.48} \\
9P/Tempel 1 & 3 555.955 & 1.51 & - & 0.094 & 0.094 & - & 1.56 & \hfill{1.46} \\
9P/Tempel 1 & 3 557.007 & 1.51 & 0.091 & 0.095 & 0.095 & 2.03 & 1.55 & \hfill{1.44} \\
9P/Tempel 1 & 3 557.955 & 1.51 & 0.084 & 0.095 & 0.095 & 1.75 & 1.54 & \hfill{1.47} \\
9P/Tempel 1 & 3 558.952 & 1.51 & 0.084 & 0.095 & 0.095 & 1.79 & 1.58 & \hfill{1.48} \\
9P/Tempel 1 & 3 559.954 & 1.51 & 0.101 & 0.101 & 0.097 & 2.57 & 1.84 & \hfill{1.60} \\
9P/Tempel 1 & 3 560.952 & 1.51 & 0.106 & 0.098 & 0.096 & 2.77 & 1.72 & \hfill{1.52} \\
9P/Tempel 1 & 3 561.953 & 1.51 & 0.087 & 0.096 & 0.095 & 1.92 & 1.63 & \hfill{1.48} \\
9P/Tempel 1 & 3 562.956 & 1.51 & 0.081 & 0.096 & 0.095 & 1.67 & 1.61 & \hfill{1.48} \\
73P-C/Schwassmann-Wachmann 3 & 3 882.367 & 0.95 & 0.104 & 0.106 & 0.107 & 2.12 & 1.31 & \hfill{1.30} \\
73P-B/Schwassmann-Wachmann 3 & 3 898.369 & 0.94 & 0.110 & 0.108 & 0.108 & 2.46 & 1.54 & \hfill{1.48} \\
8P/Tuttle & 4 481.021 & 1.04 & 0.091 & 0.093 & 0.091 & 2.16 & 1.55 & \hfill{1.36} \\
8P/Tuttle & 4 493.018 & 1.03 & 0.093 & 0.093 & 0.092 & 2.21 & 1.55 & \hfill{1.41} \\
8P/Tuttle & 4 500.017 & 1.03 & 0.095 & 0.093 & 0.091 & 2.28 & 1.47 & \hfill{1.33}  \\
103P/Hartley 2 & 5 505.288 & 1.06 & 0.095 & 0.091 & 0.097 & 2.30 & 1.35 & \hfill{1.29} \\
103P/Hartley 2 & 5 505.328 & 1.06 & 0.090 & 0.082 & 0.081 & 2.31 & 1.26 & \hfill{1.25} \\
103P/Hartley 2 & 5 510.287 & 1.07 & 0.092 & 0.081 & 0.083 & 2.31 & 1.26 & \hfill{1.06} \\
103P/Hartley 2 & 5 510.328 & 1.07 & 0.092 & 0.081 & 0.079 & 2.24 & 1.27 & \hfill{1.17} \\
103P/Hartley 2 & 5 511.363 & 1.08 & 0.090 & 0.081 & 0.081 & 2.19 & 1.23 & \hfill{1.18} \\
C/2009 P1 (Garradd) & 5 692.383 & 3.25 & 0.095 & 0.091 & 0.097 & 2.28 & 1.67 & \hfill{1.87} \\
C/2009 P1 (Garradd) & 5 727.322 & 2.90 & 0.099 & 0.083 & 0.089 & 2.54 & 1.27 & \hfill{1.49} \\
C/2009 P1 (Garradd) & 5 767.278 & 2.52 & 0.090 & 0.083 & 0.084 & 2.19 & 1.30 & \hfill{1.31} \\ 
C/2009 P1 (Garradd) & 5 813.991 & 2.09 & 0.090 & 0.085 & 0.087 & 2.18 & 1.25 & \hfill{1.33} \\
C/2009 P1 (Garradd) & 5 814.974 & 2.08 & 0.089 & 0.086 & 0.087 & 2.16 & 1.33 & \hfill{1.33} \\
C/2009 P1 (Garradd) & 5 815.982 & 2.07 & 0.090 & 0.087 & 0.089 & 2.21 & 1.57 & \hfill{1.58} \\
\hline
\end{supertabular}
\end{center}
\end{table*}
\normalsize

\begin{table}
\centering
\begin{tabular}{l l l l l}
\hline
\hline
Comet & N & \multicolumn{3}{c}{FWHM$_{\rm{intrinsic}}$~(km~s$^{-1}$)}  \\
 & & 5577.339~\r{A} & 6300.304~\r{A} &  \hfill{6363.776~\r{A}} \\
 \hline
 \hline
C/2002 V1 & 5 & 1.48 $\pm$ 0.07 & 1.43 $\pm$ 0.04 & 2.22 $\pm$ 0.32 \\
C/2002 X5 & 3 & 1.48 $\pm$ 0.07 & 1.43 $\pm$ 0.04 & 2.22 $\pm$ 0.32 \\
C/2002 Y1 & 4 & 1.57 $\pm$ 0.04 & 1.53 $\pm$ 0.06 & 2.94 $\pm$ 0.15 \\
C/2001 Q4 & 4 & 2.57 $\pm$ 0.16 & 2.49 $\pm$ 0.15 & 2.46 $\pm$ 0.13 \\
C/2002 T7 & 3& 1.77 $\pm$ 0.09 & 1.72 $\pm$ 0.07 & 2.47 $\pm$ 0.64 \\
C/2003 K4 & 3 & 1.82 $\pm$ 0.30 & 1.71 $\pm$ 0.21 & 2.55 $\pm$ 0.19 \\
9P & 10 & 1.62 $\pm$ 0.09 & 1.48 $\pm$ 0.05 & 2.23 $\pm$ 0.47 \\
73P-C & 1 & 1.27 & 1.27 & 2.13 \\
73P-B & 1 & 1.54  & 1.47 & 2.56 \\
8P & 3 & 1.53 $\pm$ 0.05 & 1.37 $\pm$ 0.04 & 2.22 $\pm$ 0.06 \\
103P & 5 & 1.28 $\pm$ 0.05 & 2.00 $\pm$ 0.07 & 2.15 $\pm$ 0.20 \\
C/2009 P1 & 2 & 1.47 $\pm$ 0.29 & 1.68 $\pm$ 0.27 & 2.41 $\pm$ 0.19 \\
C/2009 P1 & 3 & 1.34 $\pm$ 0.17 & 1.42 $\pm$ 0.15 & 2.18 $\pm$ 0.02 \\
\hline
\end{tabular}
\caption{\label{moyenne2} Average values of the intrinsic line width for each comet. The errors listed are the standard deviation of the N spectra available for each comet. 73P-C/SW3 and 73P-C/SW3 have no error quoted because we obtained only one spectrum for these comets.}
\end{table}

\begin{figure}[h]
\centerline{\includegraphics[width=\columnwidth]{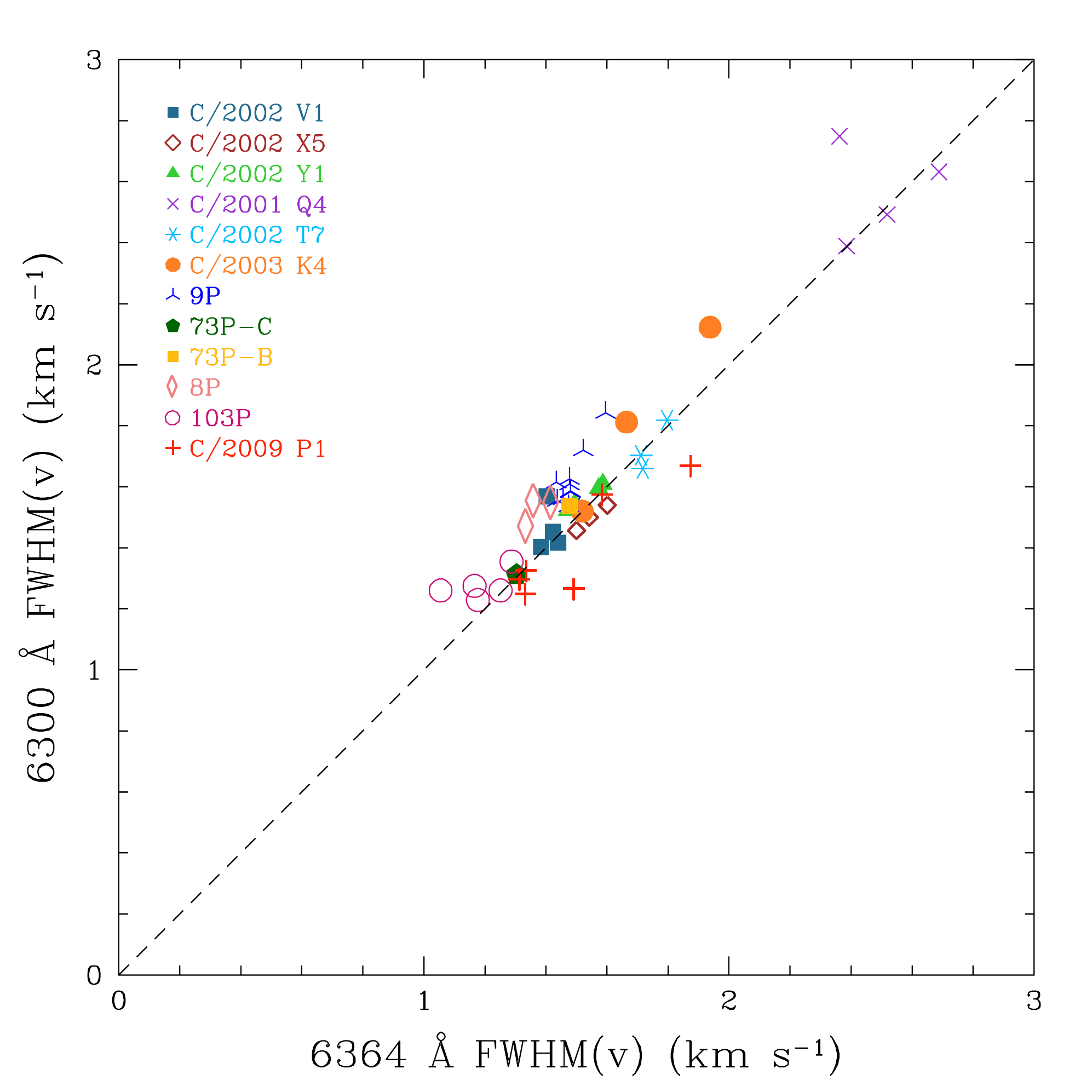}}
\caption{The 6300.304~$\r{A}$ FWHM(v) against the 6363.776~$\r{A}$ one. The different spectra for a given comet are denoted with the same symbol. Note the larger width for C/2001 Q4 (NEAT).}
\label{redwidth}
\end{figure}

\begin{figure}[h]
\centerline{\includegraphics[width=\columnwidth]{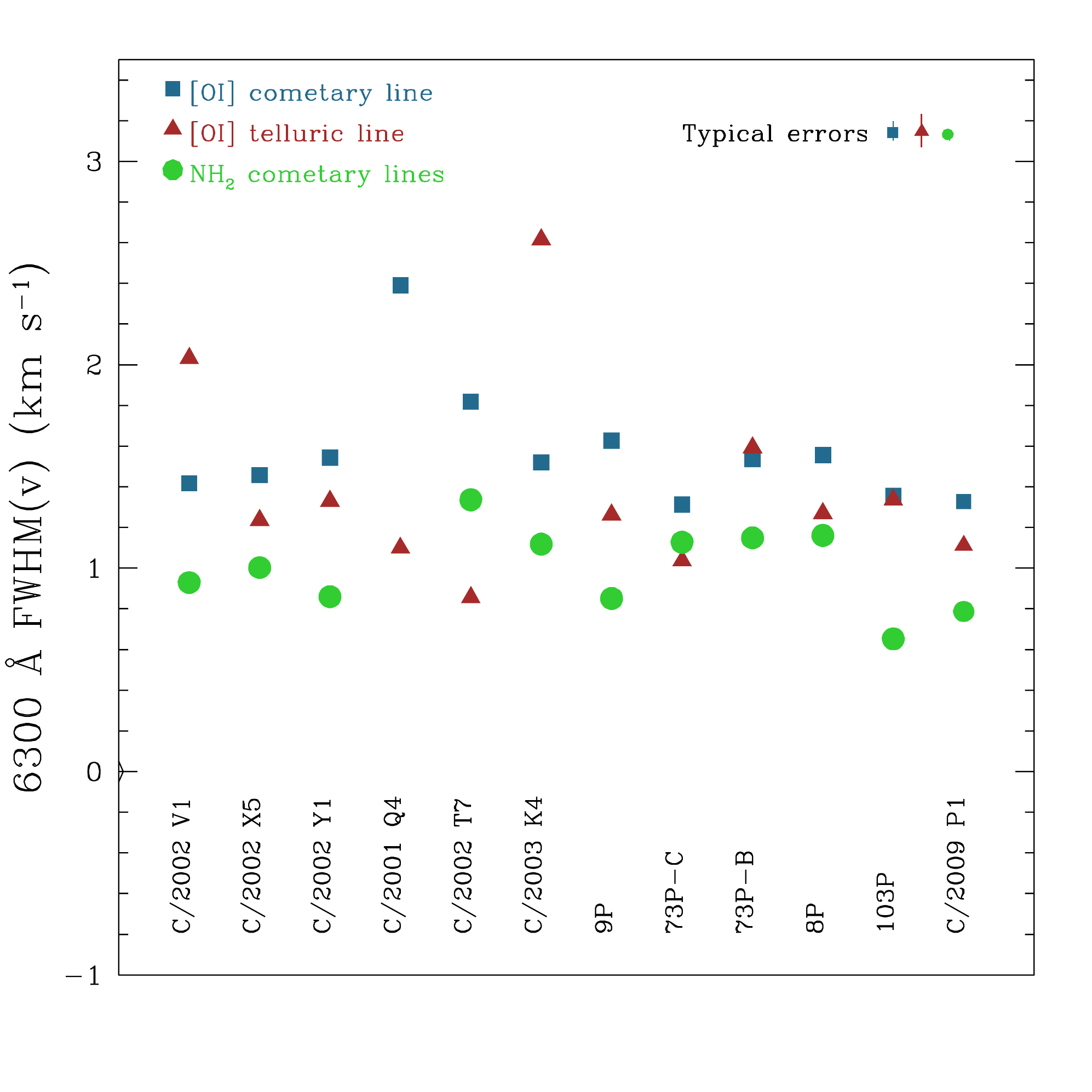}}
\caption{The FWHM(v) of the 6300.304 $\AA$~[OI] cometary line ({\footnotesize $\blacksquare$}), the [OI] telluric line ($\blacktriangle$) and nearby NH$_{2}$ cometary lines ({\scriptsize $\CIRCLE$})  (km~s$^{-1}$) for one spectrum per comet. The typical errors correspond to the standard deviation ($\sigma$) obtained considering all the spectra of comet 8P/Tuttle. We have also checked this error by measuring randomly $\sigma$ for a few spectra of other comets. For C/2001 Q4, the NH$_{2}$ cometary lines are missing because no fluorescence lines were detected at such large heliocentric distance.}
\label{cometaryline1}
\end{figure}

\begin{figure}[h]
\centerline{\includegraphics[width=\columnwidth]{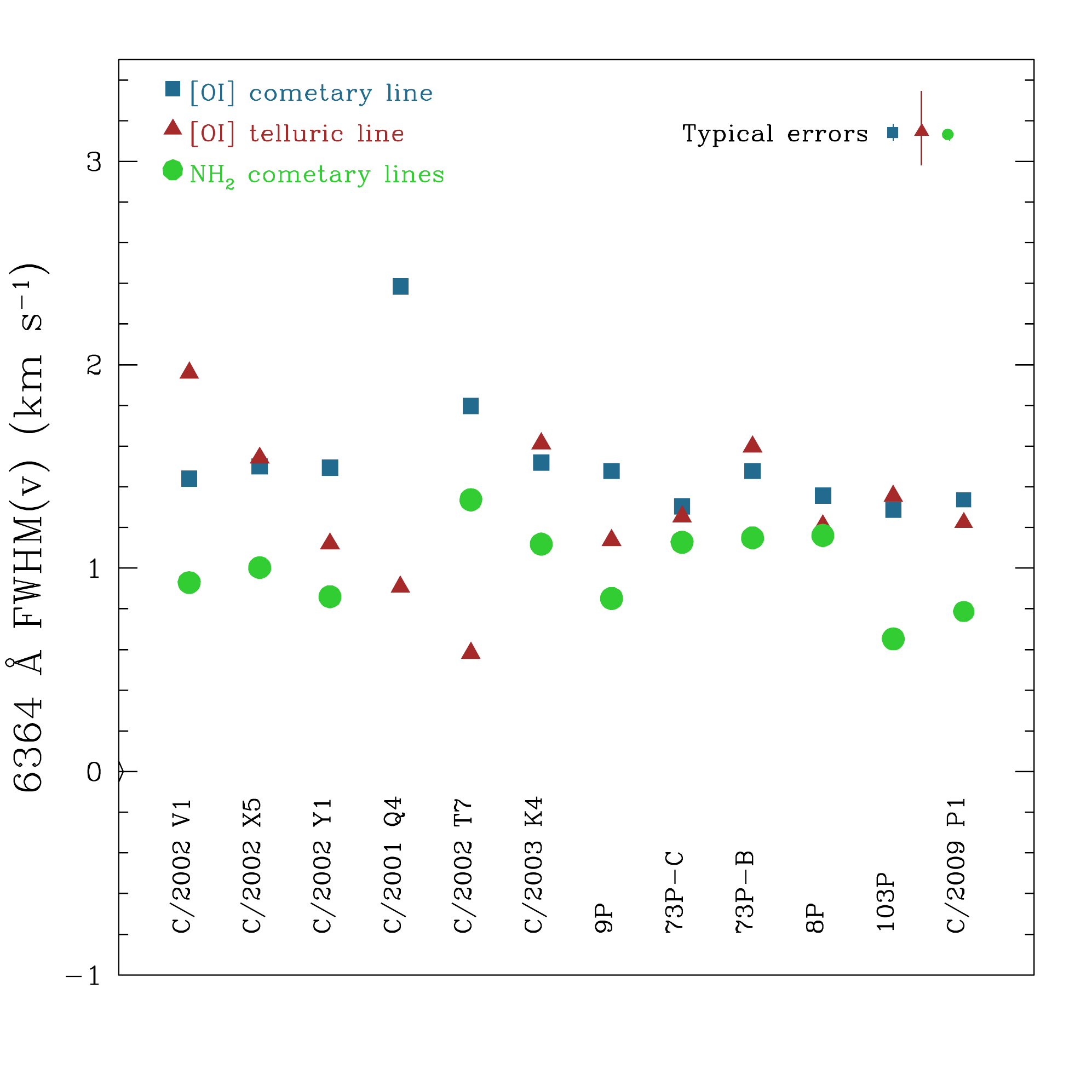}}
\caption{Same figure as Fig.~\ref{cometaryline1} for the 6363.776 $\AA$ line.}
\label{cometaryline2}
\end{figure}

\begin{figure}[h]
\centerline{\includegraphics[width=\columnwidth]{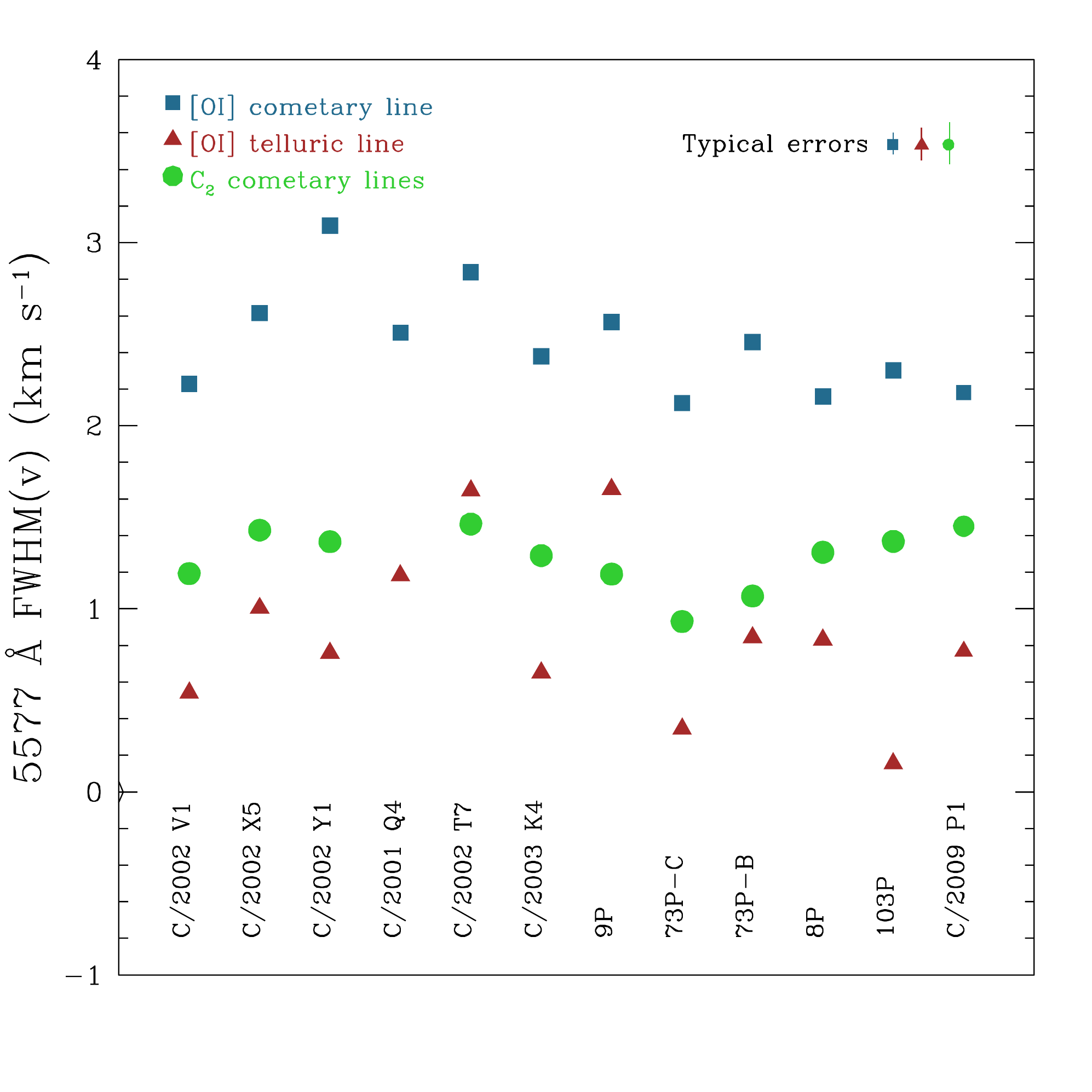}}
\caption{Same figure as Fig.~\ref{cometaryline1} for the 5577.339 $\AA$ line and with C$_{2}$ cometary lines. The larger width, by about 1~km~s$^{-1}$, of the [OI] cometary green line is obvious. It is similar for all comets, C/2001 Q4 included.}
\label{cometaryline3}
\end{figure}

\begin{figure}[h]
\centerline{\includegraphics[width=\columnwidth]{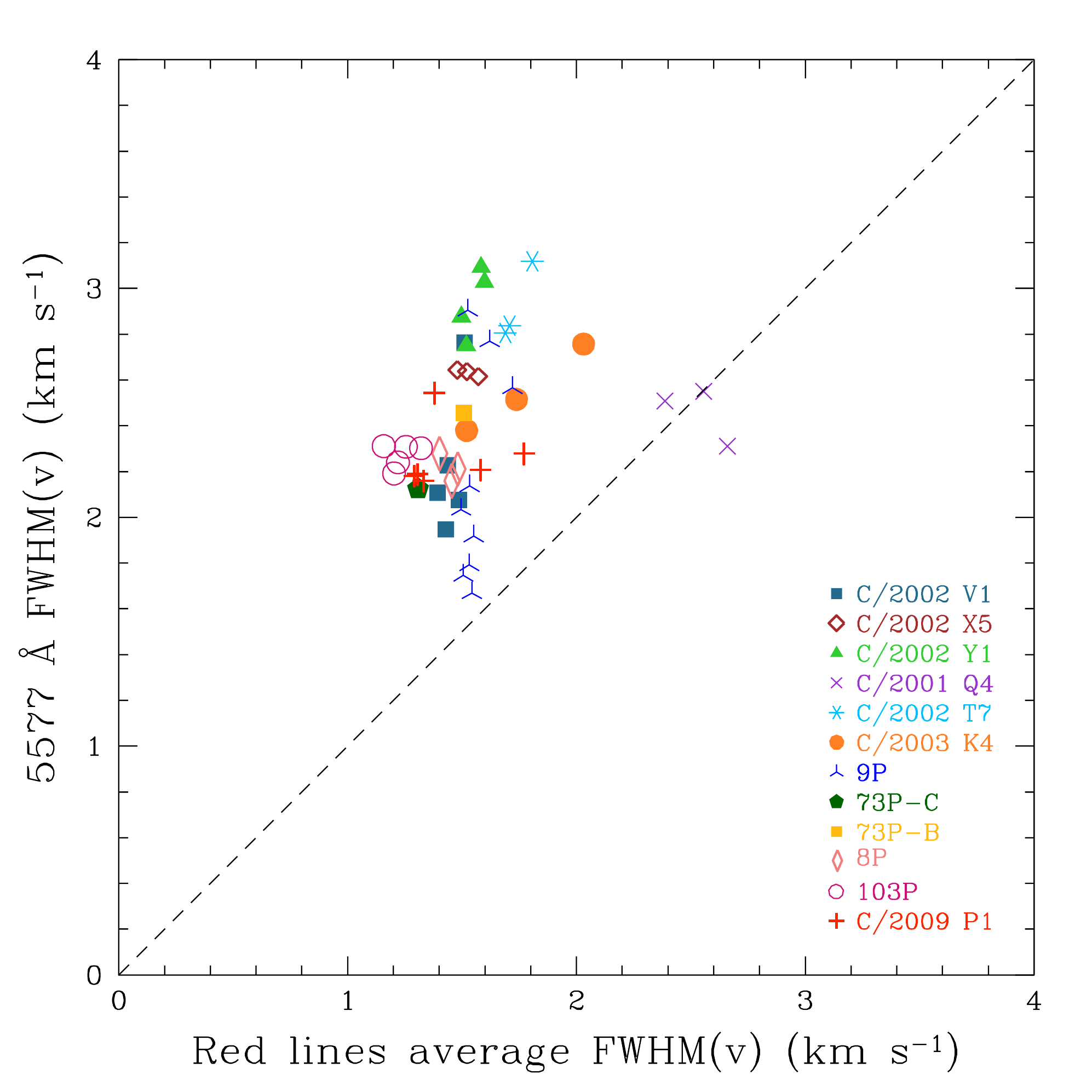}}
\caption{The 5577.339~$\r{A}$ line width against the average width of the red lines. The different spectra of a given comet are denoted with the same symbol. Note that the green line is always larger than the red lines except for C/2001 Q4.}
\label{redgreenwidth}
\end{figure}

Table~\ref{resultat2} lists the intrinsic line velocity widths (FWHM$(\rm{v})~(km~s^{-1})$). The FWHM($\rm{v}$) is obtained from the FWHM$_{\rm{intrinsic}}$($\rm{\lambda}$) given in the Eq.~\ref{fwhm} using the relation :
\begin{eqnarray}
\rm{FWHM(v)} = \frac{FWHM_{intrinsic}~(\rm{\lambda})~~Ê\textnormal{$c$}}{\lambda_{[OI]}~2~\sqrt{\ln~2}}~,
\label{width}
\end{eqnarray}

where $\lambda_{[OI]}$ corresponds to the wavelength of the considered oxygen line and $c$ is the speed of light (km~s$^{-1}$). The error on the [OI] line widths of each comet is given by the standard deviation of the N spectra (cf. Table~\ref{moyenne2}).
 This error is about $5\%$ for the red lines and $10\%$ for the green line which is fainter. The two red lines widths are equal within the errors (1.61 $\pm$ 0.34~km~s$^{-1}$ for the $6300~\AA$ line and 1.56~$\pm$~0.54~km~s$^{-1}$ for the $6364~\AA$ line) which is not a surprise since both lines are transitions from the O($^{1}$D) state to the ground state (Fig.~\ref{redwidth}). This result is also consistent with the values of \cite{Cochran2008} (1.22 $\pm$ 0.36~km~s$^{-1}$). Our mean value of the [OI] cometary green line is wider than the red lines and equal to 2.44~$\pm$~0.28~km~s$^{-1}$. This peculiarity was already noticed by \cite{Cochran2008} who found a mean velocity of 2.49~$\pm$~0.36~km~s$^{-1}$ in good agreement with our value. Figs.~\ref{cometaryline1}, \ref{cometaryline2} and \ref{cometaryline3} present the width of the three [OI] cometary lines, the three [OI] telluric lines and the width of some representative neighboring cometary lines for comparison. NH$_{2}$ lines in the red region and C$_{2}$ lines in the green region. The intrinsic average widths of the cometary NH$_{2}$ and C$_{2}$ lines are respectively $1.00 \pm 0.20$ km~s$^{-1}$ and $1.28 \pm 0.17$ km~s$^{-1}$ and correspond well to what is expected for the gas expansion velocity in the coma at 1~AU.
 The ejection velocity due to the extra energy coming from the photo-dissociation of the parent molecule should be represented as the excess velocity resulting from the subtraction of the NH$_{2}$ or C$_{2}$ cometary lines to the [OI] lines. Considering this, we find an average ejection velocities equal to 0.48~$\pm$~0.16 km~s$^{-1}$ for the red lines and 1.17~$\pm$~0.29 km~s$^{-1}$ for the green line. This analysis and Fig.~\ref{redgreenwidth} show clearly that the [OI] cometary green line is wider than the red lines. This width could be explained if the excess energy for the formation of the O($^{1}$S) state is larger than for the O($^{1}$D) state. Contrary to what is observed, theoretical models using Ly-$\alpha$ photons as excitation source give an excess velocity of 1.6 km~~s$^{-1}$ for the O($^{1}$S) state and a value of 1.8 km~~s$^{-1}$ for O($^{1}$D) in the case of water photo-dissociation \citep{Festou1981a}. \\

\begin{figure}[h]
\centerline{\includegraphics[width=\columnwidth]{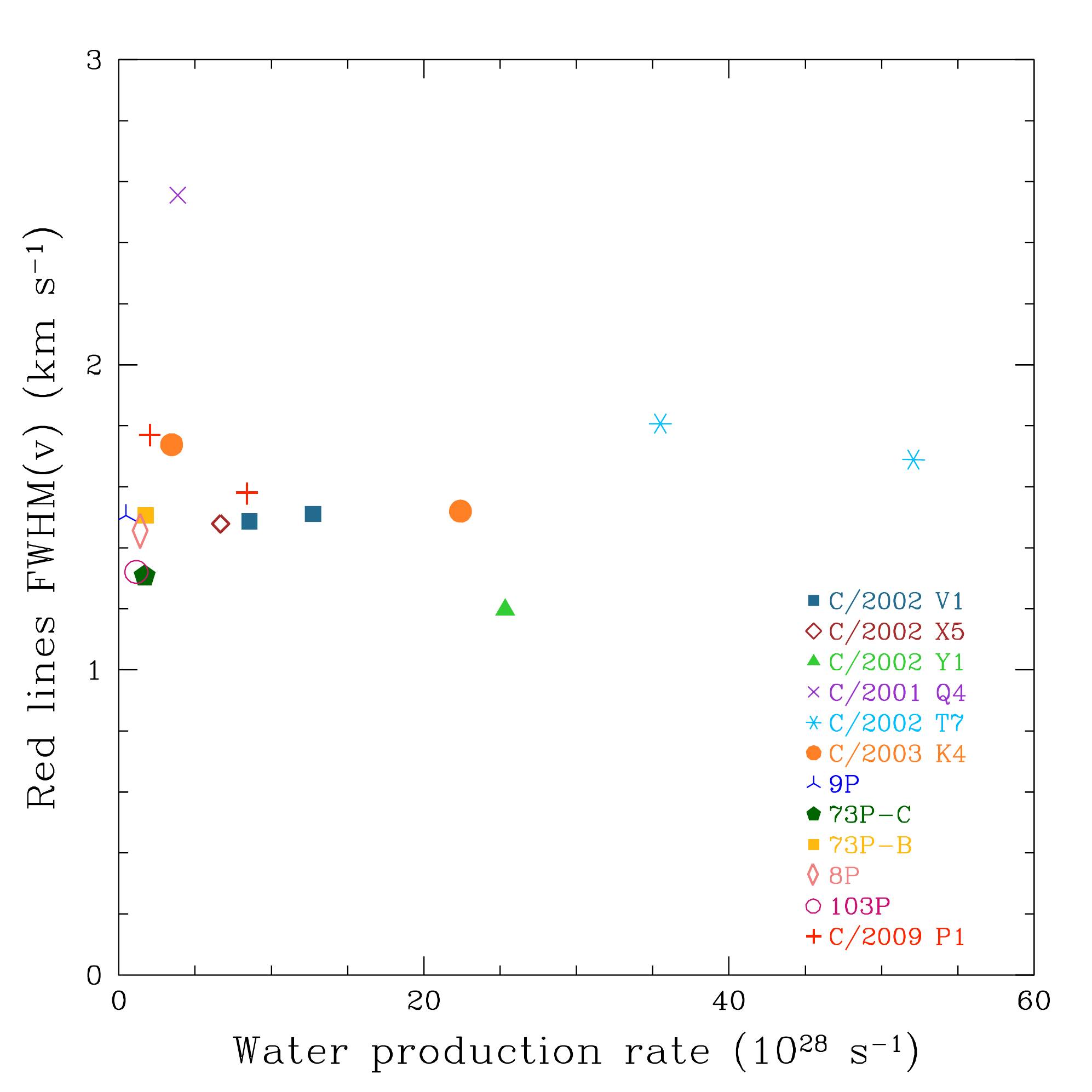}}
\caption{Average of the [OI] red lines FWHM versus the water production rate.}
\label{productionrate1}
\end{figure}


\begin{figure}[h]
\centerline{\includegraphics[width=\columnwidth]{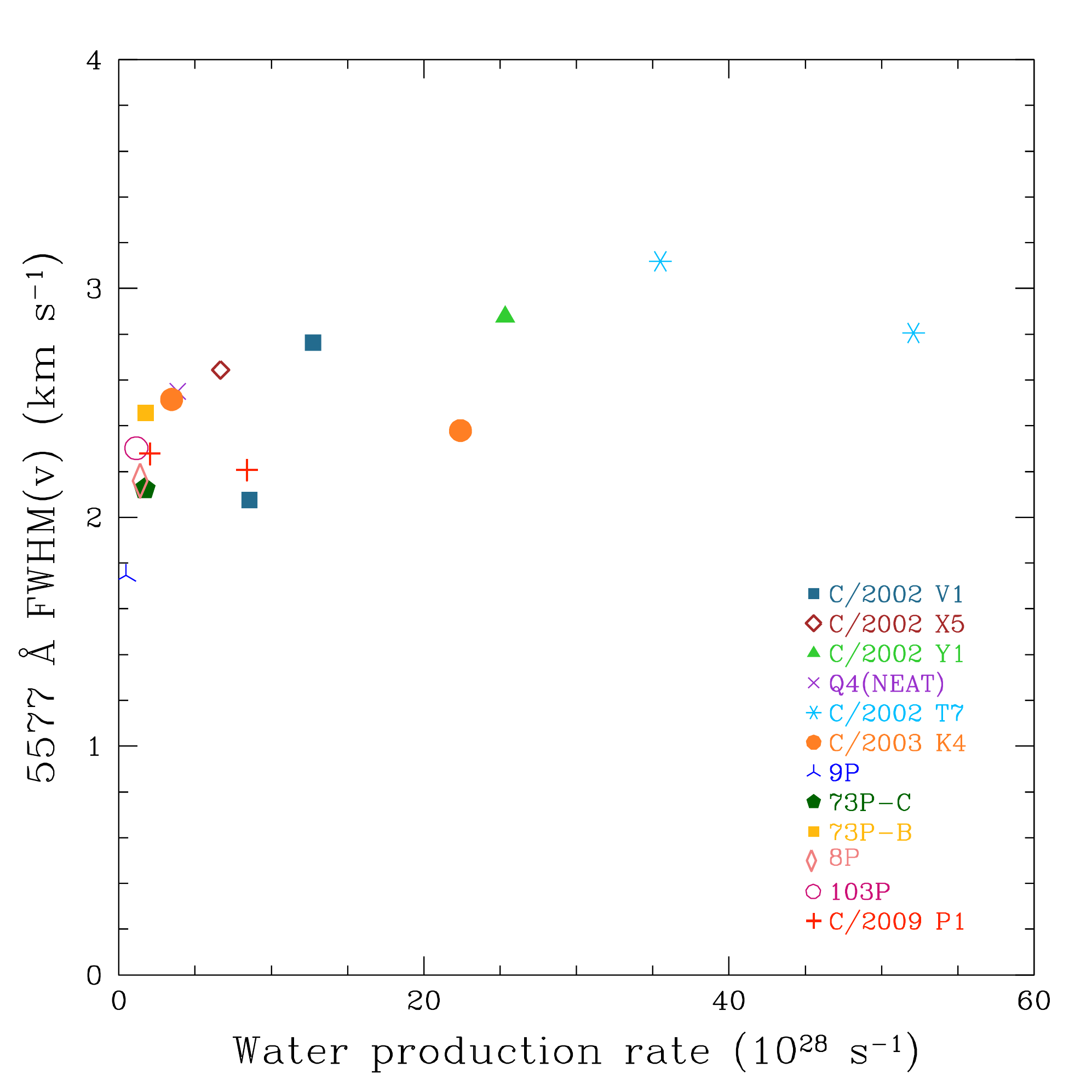}}
\caption{Same figure as Fig.~\ref{productionrate1} but for the 5577.339~$\r{A}$ line.}
\label{productionrate3}
\end{figure}

\begin{figure}[h]
\centerline{\includegraphics[width=\columnwidth]{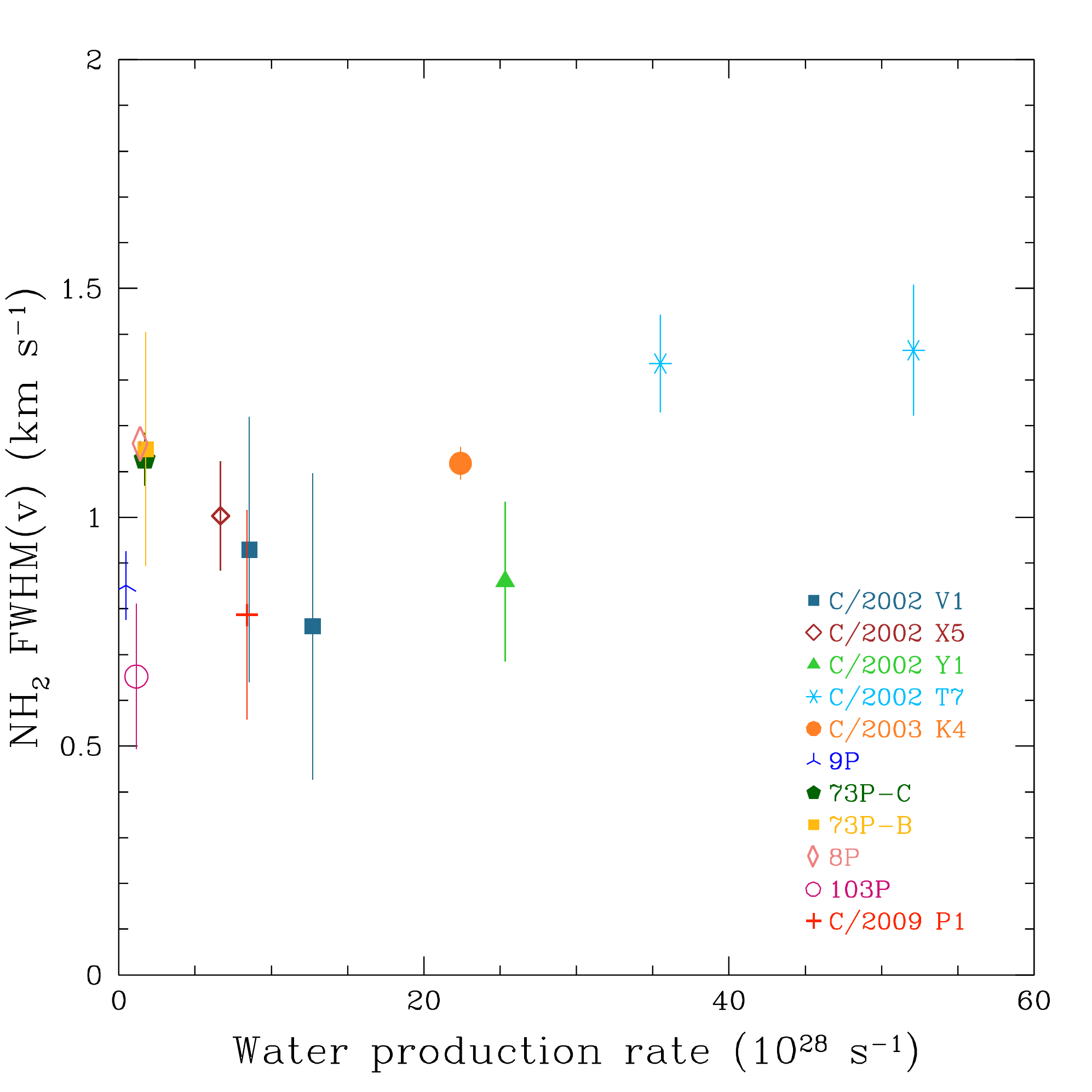}}
\caption{Same figure as Fig.~\ref{productionrate1} but for the NH$_{2}$ line.}
\label{productionrate_nh2}
\end{figure}

\begin{figure}[h]
\centerline{\includegraphics[width=\columnwidth]{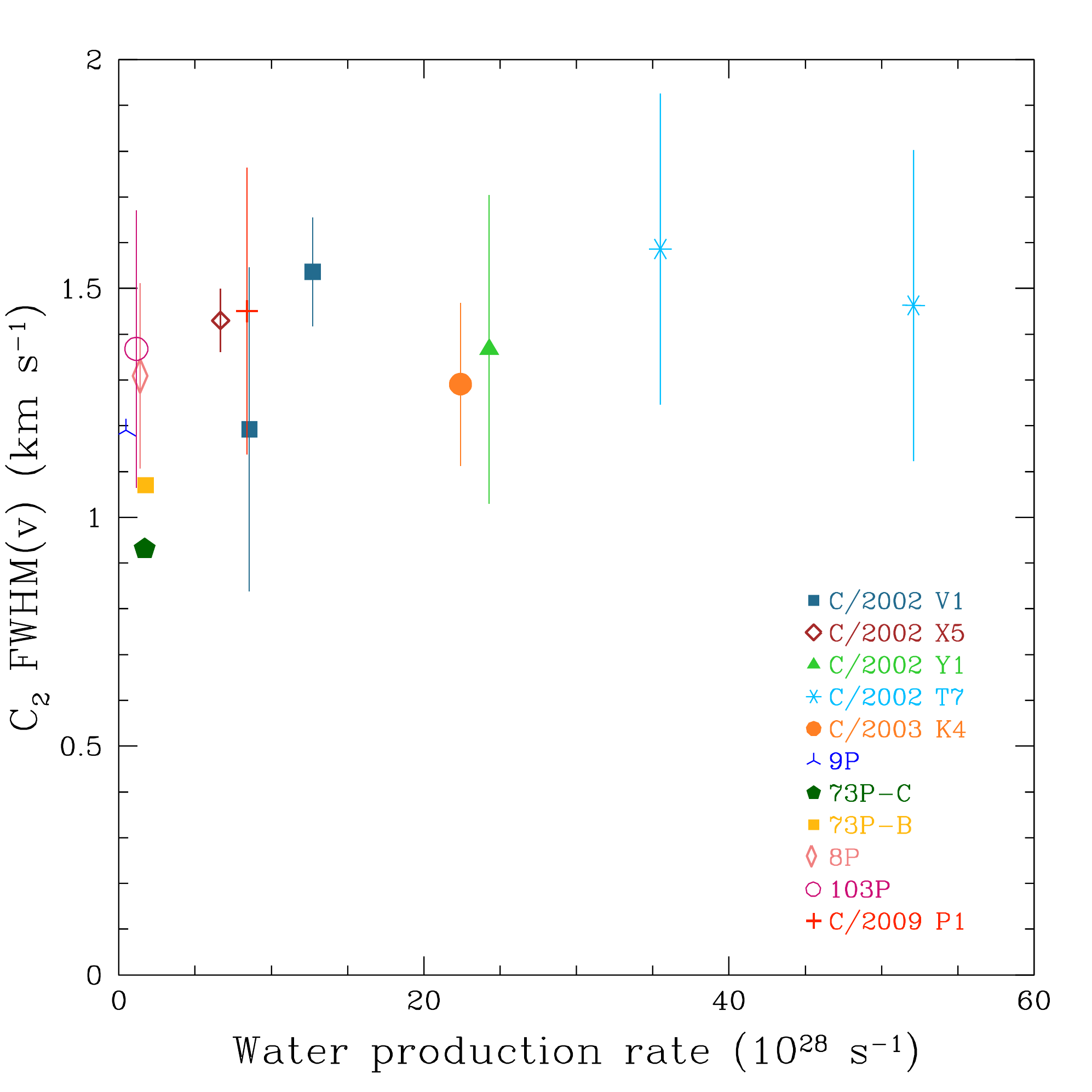}}
\caption{Same figure as Fig.~\ref{productionrate1} but for the C$_{2}$ line.}
\label{productionrate_c2}
\end{figure}

\begin{table*}[htbp]
\begin{center}
\tiny
\tablefirsthead{
\hline
\hline
Comet & $r$ &  Spectra observed date & $Q_{H_{2}O}$ ($10^{28}$ s$^{-1}$) & Date &  \hfill{Reference for $Q_{H_{2}O}$} \\
\hline
}
\tablelasttail{\hline}
\bottomcaption{\label{waterp} Water production rates  and their references. We could find $Q_{H_{2}O}$ data measured very close to our observing dates. \newline
$^{(1)}$ These water production rates were evaluated using the Jorda relation \citep{Jorda1991} with the magnitudes given by \cite{Manfroid2005}.}
\begin{supertabular}{p{4cm} >{\centering\arraybackslash}p{1cm} >{\centering\arraybackslash}p{2.2cm} >{\centering\arraybackslash}p{2cm}>{\centering\arraybackslash}p{1.8cm}p{3.2cm}}
C/2002 V1 (NEAT) & 1.22 & 2003 Jan 8 & 8.56 & 2003 Jan 8 & \hfill{\cite{Combi2011}}  \\
 & 1.01 & 2003 Mar 21 & 12.73 & 2003 Mar 21 & \hfill{\cite{Combi2011}}  \\
 C/2002 X5 (Kudo-Fujikawa) & 1.06 & 2003 Feb 19 & 6.67 & 2003 Feb 19 &  \hfill{\cite{Combi2011}} \\
 C/2002 T7 (LINEAR) & 0.68 & 2004 May 6 & 35.5 & 2004 May 5 &  \hfill{\cite{Disanti2006}} \\
  & 0.94 & 2004 May 27 & 52.1 & 2004 May 27 &  \hfill{\cite{Combi2009}} \\
  C/2003 K4 (LINEAR) & 2.61 & 2004 May 6 & 3.47 & 2004 May 6 & \hfill{$^{(1)}$} \\
   & 1.20 & 2004 Nov 20 & 22.4 & 2004 Nov 20 & \hfill{$^{(1)}$} \\
  9P/Tempel 1 & 1.51 & 2005 Jul 3 & 0.47 & 2005 Jul 3 & \hfill{\cite{Gicquel2012}} \\
  73P-C/Schwassmann-Wachmann 3 & 095 & 2006 May 27 & 1.70 & 2006 May 17 & \hfill{\cite{Schleicher2011}} \\
  73P-B/Schwassmann-Wachmann 3 & 0.94 & 2006 Jun 12 & 1.76 & 2006 May 18 & \hfill{\cite{Schleicher2011}} \\
  8P/Tuttle & 1.04 & 2008 Jan 16 & 1.4 & 2008 Jan 3 & \hfill{\cite{Barber2009}} \\
  103P/Hartley 2 & 1.06 & 2010 Nov 5 & 1.15 & 2010 Nov 31& \hfill{\cite{Knight2012}} \\
  C/2009 P1 (Garradd) & 2.07 & 2011 Sep 12 & 8.4 & 2011 Sep 17-21 & \hfill{\cite{Paganini2012}} \\
 \hline
\end{supertabular}
\end{center}
\end{table*}
\normalsize

In Figs.~\ref{productionrate1} and \ref{productionrate3}, we compare the velocity of the three oxygen lines and the H$_{2}$O production rate from litterature (cf. Table~\ref{waterp}). We find that the velocity of the green oxygen line slightly increases with the water production rate. In Fig.~\ref{productionrate_nh2} and \ref{productionrate_c2}, we provide similar plots for the velocities of C$_{2}$ and NH$_{2}$. The results are in agreement with \cite{Tseng2007} but the error bars are large because the resolution of the spectrometer is at the limit to measure the line broadening and we thus cannot improve their relation.

\section{Discussion}

\subsection{The G/R ratio at large heliocentric distance}

\begin{figure}[h]
\centerline{\includegraphics[width=\columnwidth]{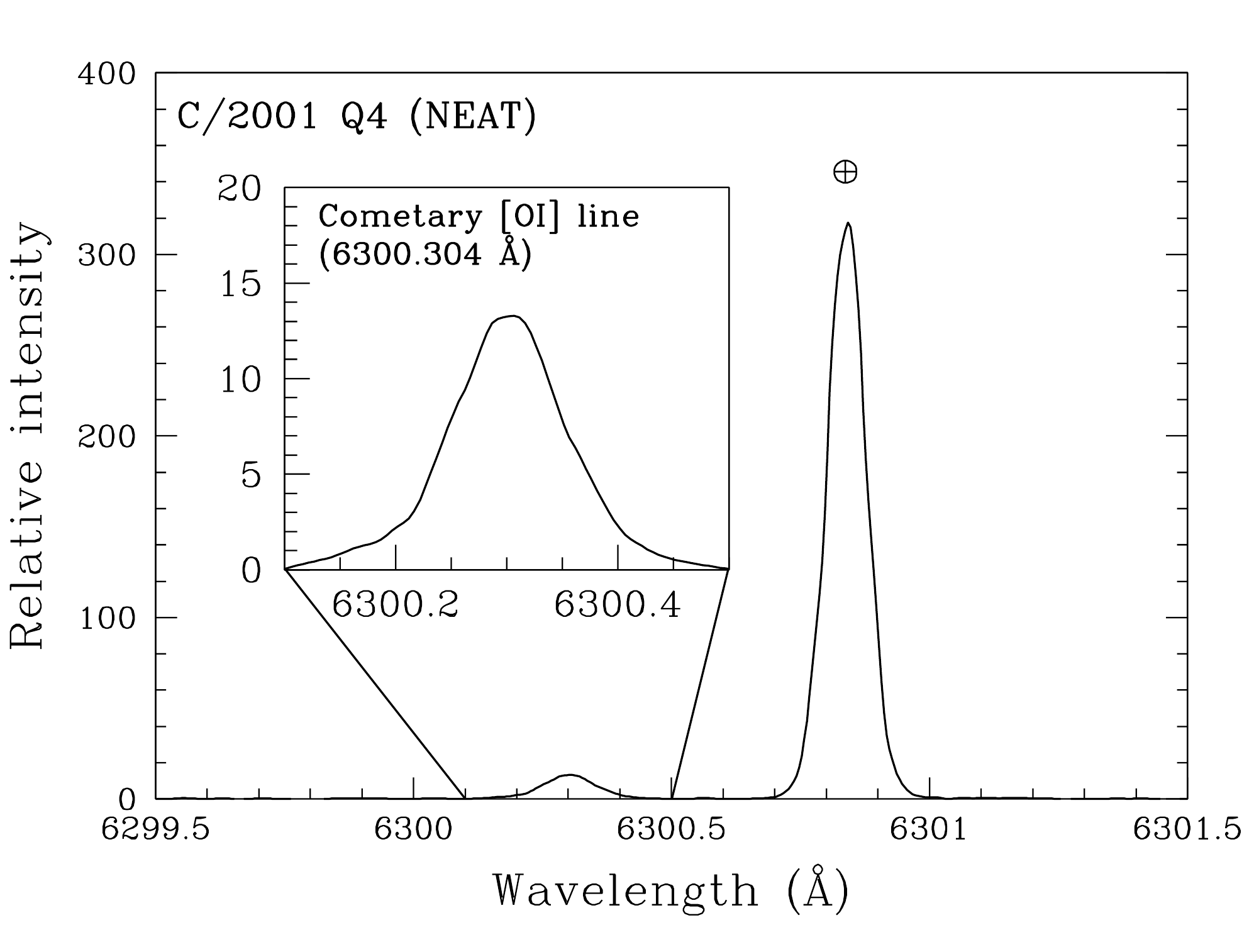}}
\centerline{\includegraphics[width=\columnwidth]{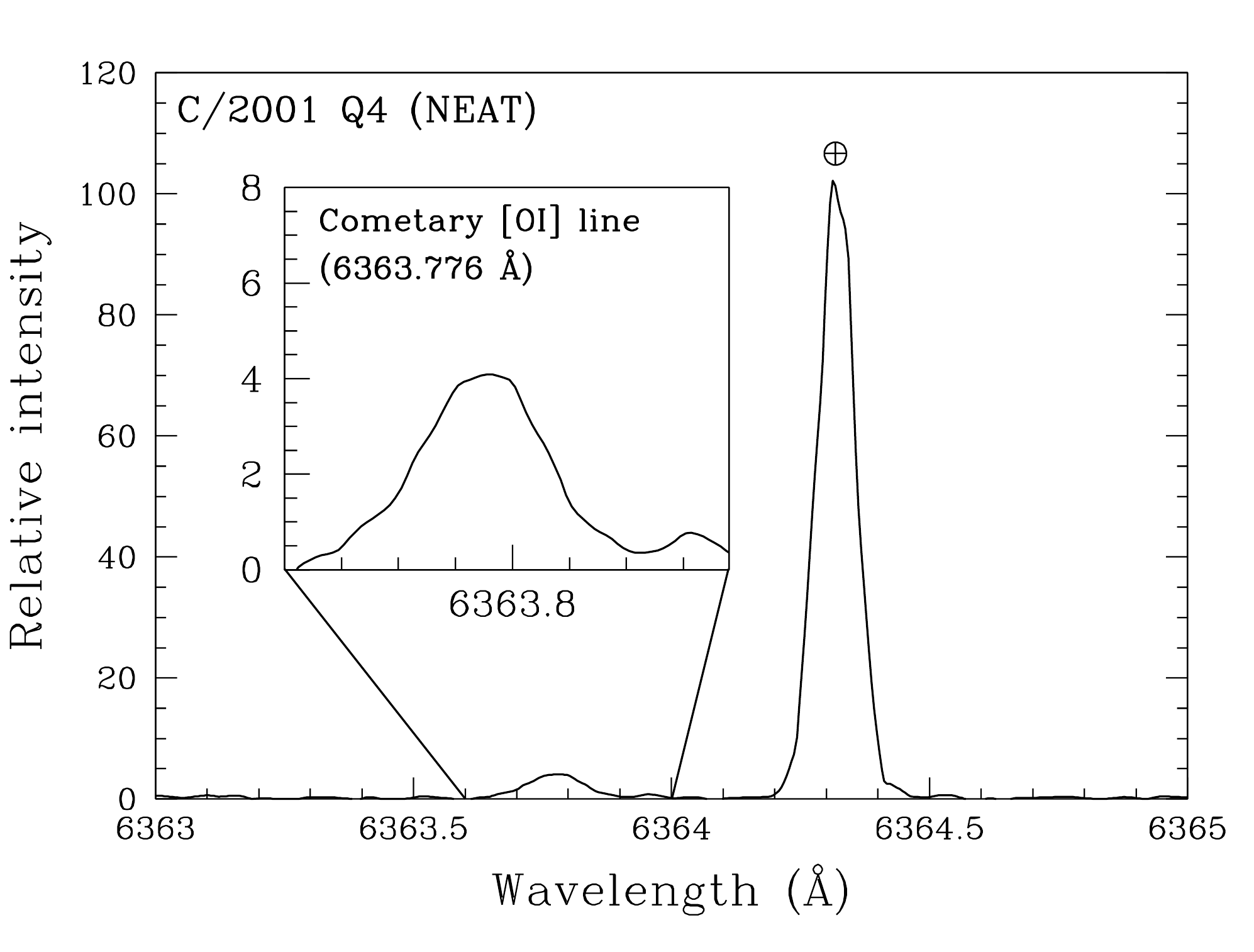}}
\centerline{\includegraphics[width=\columnwidth]{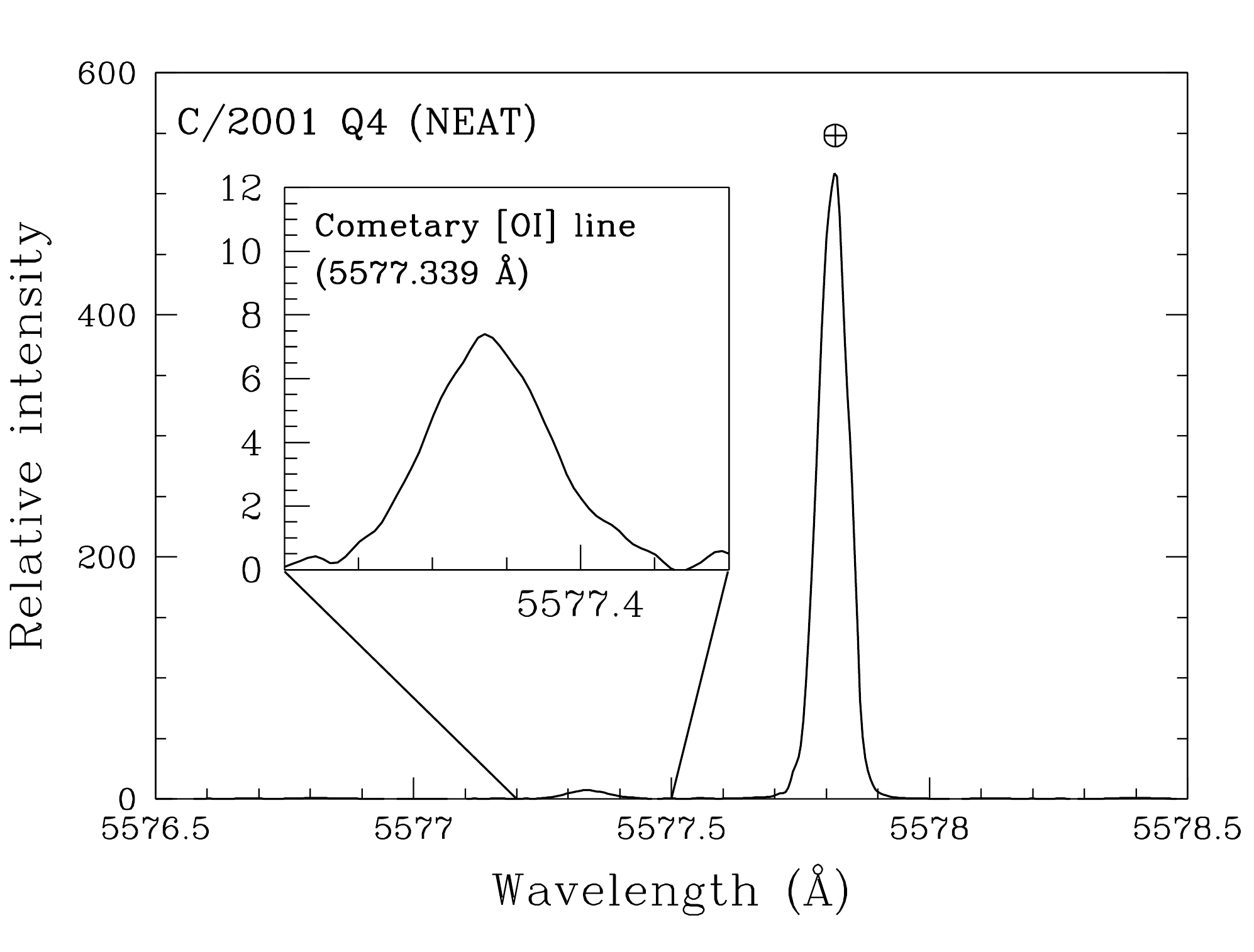}}
\caption{These two hours spectra of comet C/2001 Q4 obtained with the ESO VLT are the most distant detection of [OI] lines in a comet (at 3.7~AU). The lines are well detected and well separated from the telluric line. The width of the green and the red lines are the same.}
\label{q4}
\end{figure}

In Fig.~\ref{greenratio}, we noticed that C/2001 Q4 (NEAT) has a G/R ratio of about 0.3 clearly much higher than other comets with values around 0.1. C/2001 Q4 was at 3.7~AU from the Sun and this peculiarity led us to look more carefully on comets observed at large heliocentric distances (Fig.~\ref{q4}) \citep{Decock2011}.

\begin{figure}[h]
\centerline{\includegraphics[width=\columnwidth]{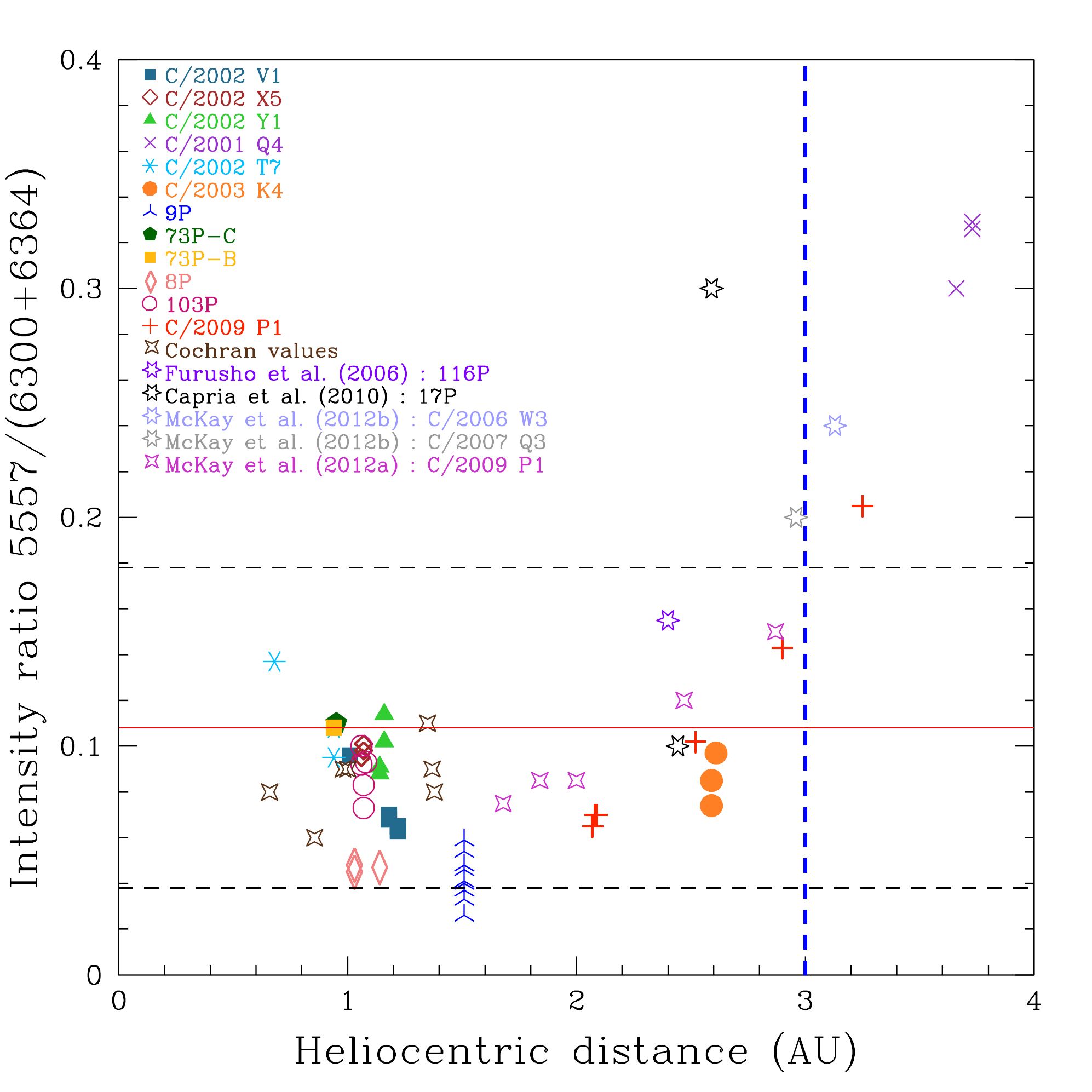}}
\caption{The G/R intensity ratio as a function of the heliocentric distance (AU). The same symbol is used for the different spectra of each comet. The most distant comets have a higher ratio. Other values from the literature are also plotted. The solid line represents the average value only for our sample of spectra and the standard deviation ($\sigma$) range is denoted with horizontal dashed lines. If we only consider the data at $r$~<~2~AU, the average value of G/R ratio is 0.09 $\pm$ 0.02. The vertical dashed line at 3 AU represents the distance out to which the sublimation of water is strongly decreasing \citep{Crovisier2000}.}
\label{garradd}
\end{figure}

If we only consider the data taken at $r$~<~2~AU, the G/R ratio average value is equal to 0.09 $\pm$ 0.02 which is in good agreement with the ratio obtained by \cite{Bhardwaj2012} and \cite{Festou1981} for H$_{2}$O as the parent molecule (see Table~\ref{bhardwaj}). The large values of G/R in Q4 (NEAT) at 3.7~AU could be explained by the increasing contribution, at large heliocentric distances, of other parent molecules producing oxygen atoms. It is well known that the sublimation of H$_{2}$O significantly decreases beyond 3~AU while the sublimation of other ices like CO and/or CO$_{2}$ dominates comet activity (i.e., \citeauthor{Crovisier2000} \citeyear{Crovisier2000}).  In order to investigate this hypothesis, we observed comet C/2009 P1 (Garradd) at four different heliocentric distances from 3.25~AU to 2.07~AU. Fig.~\ref{garradd} presents the G/R intensity ratios as a function of $r$ for all spectra. A relation between the heliocentric distance and the G/R intensity ratio can be seen : the latter is getting rapidly larger when the comet is getting farther from the Sun. We added in this plot the \cite{Cochran2001} and \cite{Cochran2008} average values coming from their sample of 8 comets. All these values are clumped around 1 AU with a ratio of $\sim$0.1. C/2001 Q4 is one of their 8 comets and its G/R ratio is $0.09~\pm~0.02$ at $r=0.98$~AU. A few other measurements found in the literature for comets observed at heliocentric distances beyond 2~AU were also added to the graph.  \cite{Furusho2006} used the Subaru telescope to analyse the forbidden oxygen lines in comet 116P/Wild 4 observed at $r~=~2.4$~AU. \cite{Capria2010} measured the intensity of these lines during and after the outburst of comet 17P/Holmes which occurred on October 24, 2007 when the comet was located at about $r$~=~2.5~AU. \cite{McKay2012} studied the [OI] lines of the two comets C/2006 W3 Christensen and C/2007 Q3 Siding Spring, respectively at 3.13 and 2.96 AU, with the ARCES echelle spectrometer of 3.5-m telescope at Apache Point Observatory.
While the quality of the data are not as good because the lines are faint and sometimes heavily blended with the telluric line, all these measurements show a relatively high value of the G/R ratio when $r$~>~3~AU. This confirms the hypothesis that oxygen is also coming from other molecules. CO and/or CO$_{2}$ are obvious candidates as they produce large values of the G/R ratio (see Table~\ref{bhardwaj}). 

Using HST/STIS\footnote{HST/STIS is the acronym for Hubble Space Telescope / Space Telescope Imaging Spectrograph (http://www.stsci.edu/hst/stis).} observations, \cite{Feldman2004} estimated at 4$\%$ the CO abundance relative to water in comet C/2001 Q4 located at 1~AU from the Sun. The study made by \cite{Biver2012} on comet C/2009 P1 provided an average value of 5$\%$ for the CO abundance when the comet was at 1.9~AU at pre-perihelion phase (October 2011). At large heliocentric distances, CO/H$_{2}$O had to be even higher. So the CO molecule could be a candidate to explain such a large G/R ratio. However, \cite{Bhardwaj2012} showed that after the photo-dissociation of H$_{2}$O, the next main source for the green line emission is CO$_{2}$ with 10-40$\%$ abundance relative to water. They compared their model with the observational values obtained for C/1996 B2 (Hyakutake) by \cite{Morrison1997} and \cite{Cochran2008} when the comet was around 1~AU. Hyakutake is also rich in CO and its abundance was evaluated at 22$\%$ at 1~AU \citep{Biver1999}. Assuming a CO$_{2}$ abundance of $1\%$ and a O($^{1}$S) yield of $0.2\%$, \cite{Bhardwaj2012} concluded that the production rates of the O($^{1}$S) are similar for CO and CO$_{2}$ ($\sim 10-25\%$) with such abundances. Therefore, since C/2001Q4 and C/2009 P1 have less CO than Hyakutake, CO$_{2}$ molecules could rather be the main contributor to the formation of the green line emission for these two comets observed at large heliocentric distance. CO$_{2}$ measurements are unfortunately very rare because they are difficult to obtain to confirm this hypothesis. But if this is the case, G/R ratios estimated might be used as a proxy for the CO$_{2}$ relative abundance but such a relation should be carefully calibrated on a sample of comets with known CO$_{2}$.

\paragraph{}

Using Eq.~\ref{intensity} and only considering the main parent species of oxygen atoms (i.e H$_{2}$O, CO$_{2}$ and CO), we can write  the G/R ratio as :

\scriptsize
\begin{eqnarray}
G/R = \frac{I^{green}}{I^{red}} = \frac{W^{green}_{H_{2}O}~~\beta^{green} ~~ N_{H_{2}O} + W^{green}_{CO_{2}}~~\beta^{green} ~~ N_{CO_{2}} + W^{green}_{CO}~~\beta^{green} ~~ N_{CO}}{W^{red}_{H_{2}O}~~\beta^{red} ~~ N_{H_{2}O} + W^{red}_{CO_{2}}~~\beta^{red} ~~ N_{CO_{2}} + W^{red}_{CO}~~\beta^{red} ~~ N_{CO}}.
\label{gsurr}
\end{eqnarray}
\normalsize

where $W^{a}_{b}=\alpha^{a}_{b}~\tau^{-1}_{b}$ corresponds to the effective production rates of O($^{1}$S) and O($^{1}$D) states\footnote{The production rates $W^{a}_{b}$ are given in Tables 1 and 2 of \cite{Bhardwaj2012} (i.e $W^{red}_{H_{2}O}=8~10^{-7}$~s$^{-1}$, $W^{green}_{H_{2}O}=6.4~10^{-8}$~s$^{-1}$, $W^{green}_{CO_{2}}=7.2~10^{-7}$~s$^{-1}$, $W^{red}_{CO_{2}}=1.2~10^{-6}$~s$^{-1}$, $W^{green}_{CO}=4~10^{-8}$~s$^{-1}$ and $W^{red}_{CO}=5.1~10^{-8}$~s$^{-1}$) and $\beta^{green}$~=~0.91 \citep{Slanger2006}.}. 
However, if we consider that oxygen atoms only come from H$_{2}$O and CO molecules, we found that G/R ratio does not change strongly : it varies from 0.08 to 0.1 for a CO abundance between 10$\%$ and 80$\%$. Therefore, we assume that the [OI] atoms are only produced by the H$_{2}$O and CO$_{2}$ photo-dissociations.
Hence, Eq.~\ref{gsurr} can be simplified as :

\begin{eqnarray}
G/R = \frac{I^{green}}{I^{red}} = \frac{W^{green}_{H_{2}O}~~\beta^{green} ~~ N_{H_{2}O} + W^{green}_{CO_{2}}~~\beta^{green} ~~ N_{CO_{2}}}{W^{red}_{H_{2}O}~~\beta^{red} ~~ N_{H_{2}O} + W^{red}_{CO_{2}}~~\beta^{red} ~~ N_{CO_{2}}}
\label{abundance2}
\end{eqnarray}

Therefore, the CO$_{2}$/H$_{2}$O abundance can be computed by :
\begin{eqnarray}
\frac{N_{CO_{2}}}{N_{H_{2}O}} = \frac{(G/R)~W^{red}_{H_{2}O} - \beta^{green}~W^{green}_{H_{2}O}}{\beta^{green}~W^{green}_{CO_{2}}-(G/R)~W^{red}_{CO_{2}}}.
\label{abundance}
\end{eqnarray}

\begin{figure}[h]
\centerline{\includegraphics[width=\columnwidth]{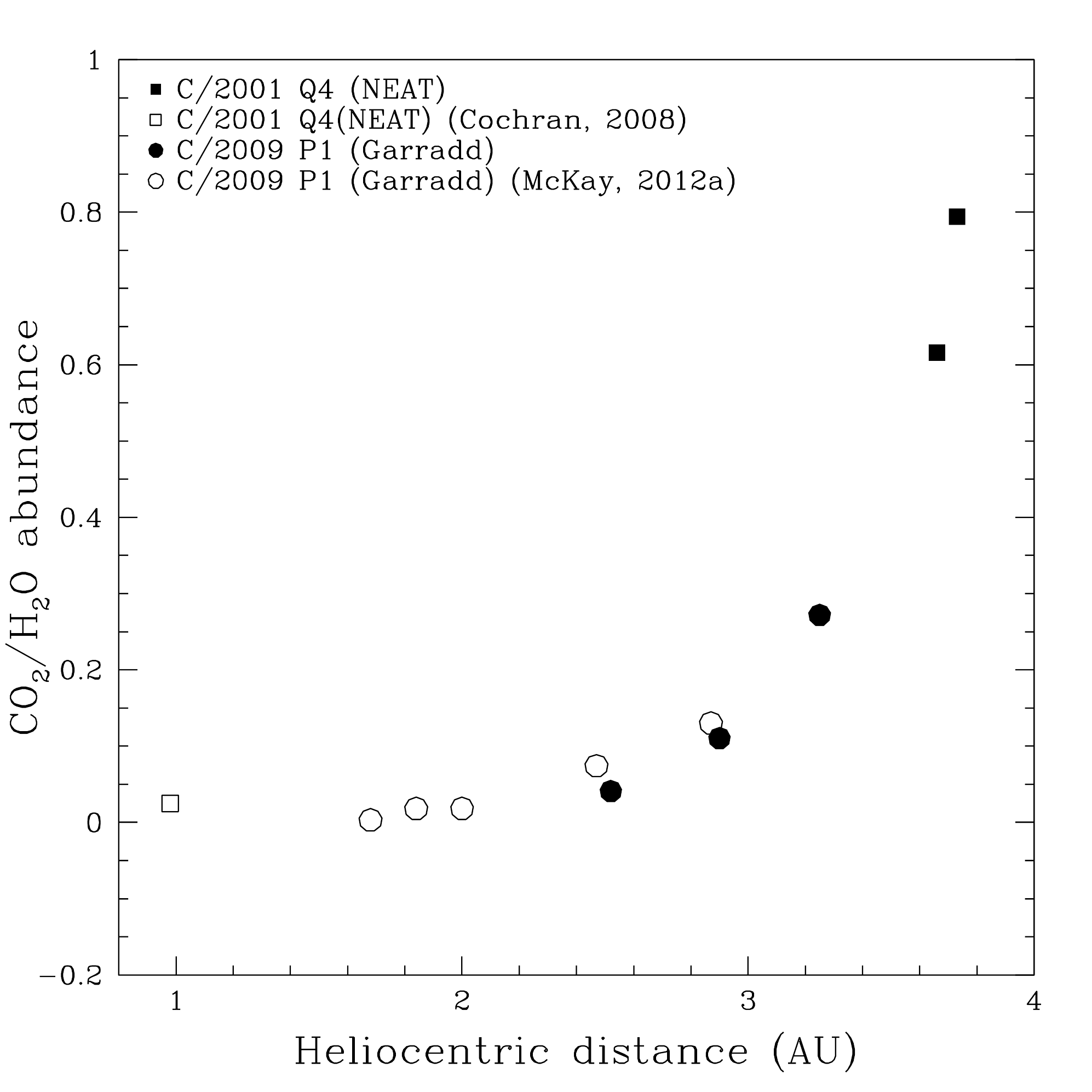}}
\caption{Evolution of the CO$_{2}$/H$_{2}$O abundance as a function of the heliocentric distance (AU). Assuming that a bigger G/R ratio comes from a more important contribution of CO$_{2}$ for large heliocentric distance comets, measuring this ratio can give us the CO$_{2}$ abundance from ground observations.}
\label{CO2abundance}
\end{figure}

The $W^{a}_{b}$ are independent of the heliocentric distance because all of them depend on the same heliocentric distance. This is also the case for the values given in Table~\ref{bhardwaj}. Considering the same values for $W^{a}_{b}$ and $\beta^{green}$, we also computed the CO$_{2}$/H$_{2}$O abundance from the G/R ratios of comets C/2001 Q4 and C/2009 P1 measured by \cite{Cochran2008} and \cite{McKay2012a}, respectively. 
These results are shown in Fig.~\ref{CO2abundance}. We noticed that  CO$_{2}$/H$_{2}$O is as high as 75$\%$ in C/2001 Q4 and that CO$_{2}$ starts to contribute to the G/R ratio at 2.5~AU in the comets. This heliocentric distance limit is also shown by \cite{Ootsubo2012} from a sample of 17 comets observed with the AKARI IR space telescope.

Therefore, the forbidden oxygen lines measurement could become a new way to determine the CO$_{2}$ abundance in comets at different heliocentric distances from ground observations while the direct measurement of CO$_{2}$ molecules is only possible from space.

\subsection{The [OI] lines widths}

One explanation for the large width of the [OI] green line could be that exciting photons producing oxygen atoms in the $^{1}$S state have more energy than the Lyman $\alpha$ photons and/or come from an other parent molecule. Indeed, Fig.~7 of \cite{Bhardwaj2012} shows that the wavelength region for the major production of O($^{1}$S) coming from CO$_{2}$ seems to be the 955-1165~\r{A} band. New theoretical studies could be necessary to evaluate the excess velocity for the O($^{1}$S) state coming from the photo-dissociations of both H$_{2}$O and CO$_{2}$. \\
In Fig.~\ref{redgreenwidth}, we also notice that the [OI] red lines of the comet C/2001 Q4 are wider than in other comets while the width of the green line is normal with respect to other comets. This peculiarity seems real as both red lines show the same width and they are 2-$\sigma$ away from the average value of the sample (1.58~$\pm$~0.30~km~s$^{-1}$). The telluric lines in C/2001 Q4 have the usual width (0.99~$\pm$~0.10~km~s$^{-1}$ for the 6300~$\AA$~line, 0.87~$\pm$ 0.04~km~s$^{-1}$ for the 6364~$\AA$~line and 0.80~$\pm$~0.26~km~s$^{-1}$ for the green line), excluding a problem with the spectra or the analysis. The width of the red lines and the green line is similar in this only case (2.57~$\pm$~0.16~km~s$^{-1}$ for the 6300~$\AA$ line, 2.49~$\pm$~0.15~km~s$^{-1}$ for the 6364~$\AA$ and 2.46~$\pm$~0.13~km~s$^{-1}$ for the 5577~$\AA$ line) and could give us clues about the process at play. 
As previously discussed, at large heliocentric distances, the [OI] lines could mainly come from the CO$_{2}$ molecules which preferentially photo-dissociate in the $^{1}$S state while at low heliocentric distances, they essentially come from H$_{2}$O which preferentially dissociates in the $^{1}$D state (Table~\ref{bhardwaj}). This means that at large distances, the red lines are mainly produced through the $^{1}$S-$^{1}$D channel together with the green line, while at low distances as H$_{2}$O dominates they are mostly produced directly from the $^{1}$D state. We thus expect the widths of the red and green lines to be similar at high distances because the O($^{1}$S) and O($^{1}$D) atoms are mostly produced from a same molecule, while they can differ at low distances, as observed. CO$_{2}$ is the best candidate and the larger width could be explained by the main excitation source of CO$_{2}$, the 955-1165~$\AA$ band, which is more energetic than Ly-$\alpha$ photons, as shown in Fig.~11 of  \cite{Bhardwaj2012}. This particularity is not seen for comet C/2009 P1 (Garradd) maybe because the spectrum is not taken at sufficiently large heliocentric distance and the CO$_{2}$ abundance was lower than in C/2001 Q4 (NEAT). Anyway, this hypothesis needs to be confirmed by observing other comets at large heliocentric distances (>~$3.5$~AU).

\subsection{103P/Hartley 2}

\begin{table}
\centering
\begin{tabular}{l| c| c|  r}
\hline
\hline
JD - 2~450~000.5 & r(AU) & G/R & $N_{CO_{2}}/N_{H_{2}O}$ \\
\hline
5 505.288 & 1.06 & 0.09 & 0.028 \\
5 505.328 & 1.06 & 0.10 & 0.040 \\
5 510.287 & 1.07 & 0.07 & 0.001 \\
5 510.328 & 1.07 & 0.08 & 0.015 \\
5 511.363 & 1.08 & 0.09 & 0.030 \\
\hline
\end{tabular}
\caption{\label{hartley}The CO$_{2}$/H$_{2}$O abundances estimated from Eq.~\ref{abundance} for the comet 103P/Hartley~2. }
\end{table}

The Jupiter Family comet 103P/Hartley 2 has been found poor in CO \citep{Weaver2011} and
rich in CO$_{2}$ with an CO$_{2}$ abundance of $\sim20\%$ relative to the water from EPOXI measurements \citep{AHearn2011} and with an abundance of $\sim10\%$ from the ISO observations \citep{Crovisier1999}. Thanks to 
Eq.~\ref{abundance}, we could evaluate the CO$_{2}$ abundances from our G/R ratios
measured at $\sim1$ AU and compare them with the
values obtained by EPOXI and ISO observations. The results are given in Table~\ref{hartley}.
The mean G/R ratio of $0.09~\pm~0.01$ is similar to those of other comets
observed below 2.5 AU and give a relative abundance of CO$_{2}$ of
only $\sim3\%$ while a G/R ratio of $\sim0.1$ to $0.2$ would be needed to
reach a CO$_{2}$ abundance as high as $\sim10\%$ to $20\%$. \\
We do not confirm a high value of CO$_{2}$ abundance for 103P/Hartley~2
as claimed by \citep{McKay2012b} or a large dispersion in the
abundances. Their values might be higher because they used a
$W^{green}_{H_{2}O}$ of \cite{Bhardwaj2012} equal to 2.6~10$^{-8}$ s$^{-1}$
while the value provided in Table 1 of this paper is 6.4~10$^{-8}$ s$^{-1}$.\\
The discrepancy with the EPOXI and ISO observations CO$_{2}$ abundance is interesting. Below 2.5 AU, G/R values 
are distributed from 0.05 for comets like 9P/Tempel~1 and 8P/Tuttle and can go
up to 0.12 for C/2002 T7~(LINEAR). The spread from comet to comet is higher than
the errors estimated from the dispersion from the spectra of a same comet
which could be explained by an intrinsic variation of G/R from comet to comet due to the content of CO$_{2}$.
The G/R range, of the order of 0.1 (implying a CO$_{2}$ variation of $\sim20\%$)
is of the same order as the range of CO$_{2}$ abundance in comets of various origins
measured below 2.5 AU \citep{AHearn2012}. 

The fact that some comets, like 9P/Tempel~1 and 8P/Tuttle, have values lower than 
the pure water case (giving a G/R$_{\rm{min}}$~=~0.08 from \cite{Bhardwaj2012}) and that the 103P/Harley~2 G/R ratio is too low 
compared to EPOXI CO$_{2}$ abundances, could 
indicate a problem with the models. The pure water ratio should be 0.05 or lower 
based on the comets with the smallest values (9P/Tempel~1 has a CO$_{2}$ abundance of $\sim7\%$
(\cite{Feaga2007a}, \cite{Feaga2007b}). \cite{Bhardwaj2012} tried $\alpha^{green}_{H_{2}O}$ values
from $0\%$ to $1\%$ and finally chose $1\%$, but this value is uncertain. Without CO$_{2}$,
G/R$_{\rm{min}}$~=~$\beta^{green}~W^{green}_{H_{2}O}~/~W^{red}_{H_{2}O}$. If we use $0.5\%$ for $\alpha^{green}_{H_{2}O}$, G/R$_{\rm{min}}$ is equal to 0.04 instead of
0.08 which would be in better agreement with our smallest values.

\subsection{Deep Impact}

\begin{figure}[h]
\centerline{\includegraphics[width=\columnwidth]{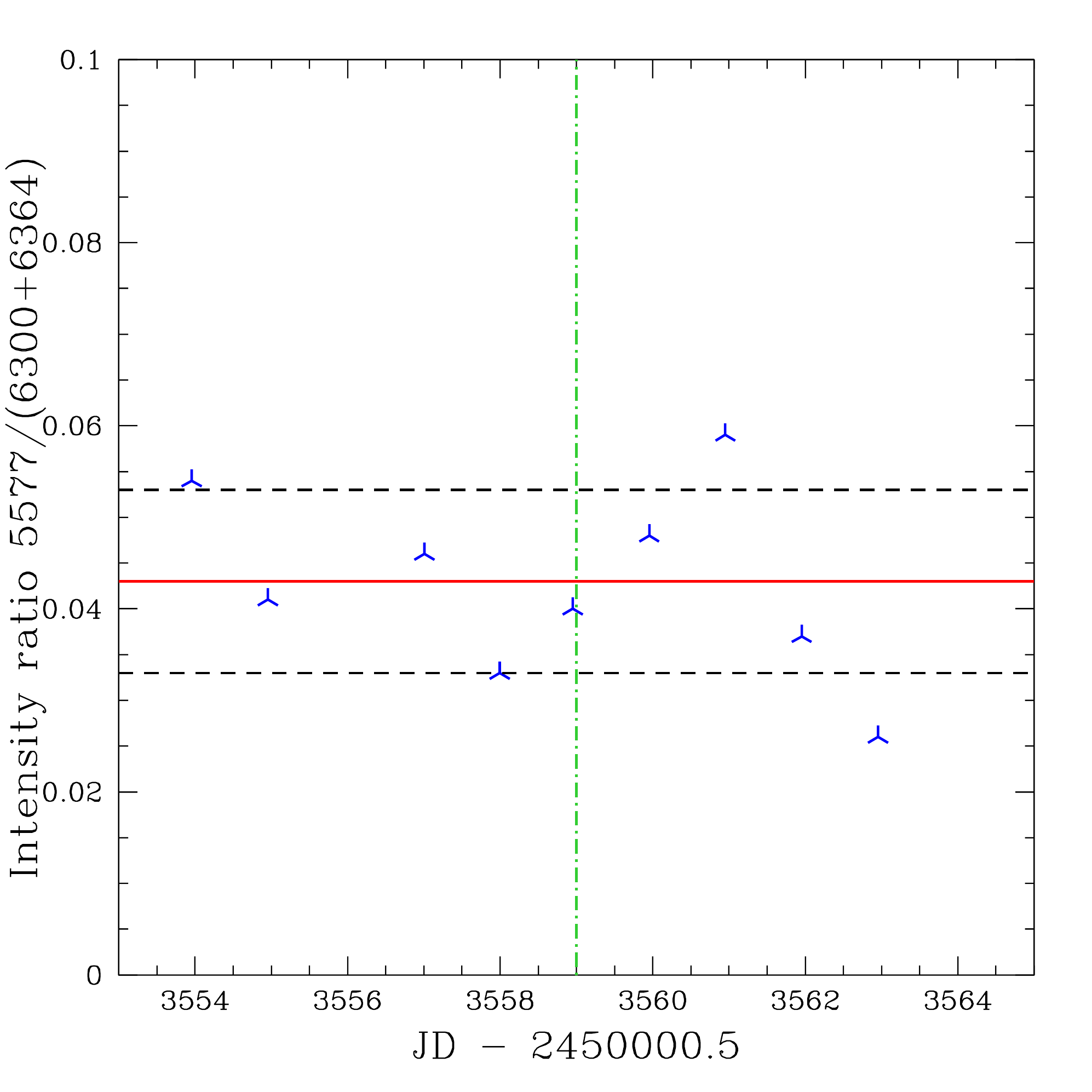}}
\caption{The G/R intensity ratio of comet 9P/Tempel 1 from JD $2453553.955$ to $2453562.956$ UT. The Deep Impact event corresponds to the vertical line (53559 UT). The average value is shown by the solid red line and the dashed lines correspond to the standard deviation ($\sigma$). The mean value around the impact (from 2453553.955 UT to 2453562.956 UT) is $0.04~\pm~0.01$. No variation of the ratio was noticed after the impact. }
\label{tempel}
\end{figure}

The spectra of comet 9P/Tempel~1 were analysed before and after the July 4 collision with the Deep Impact spacecraft. No difference was observed in the intensity ratios and the lines widths before and after the impact (Fig.~\ref{tempel}). We have no spectrum during the impact. The first spectrum after the impact was taken at 2453559.954 UT (i.e. $\sim$6 hours after the impact). \cite{Cochran2008} did not notice any change during the impact either. 

\subsection{Solar activity}

Our data were obtained over the 23$\rm{^{th}}$ solar cycle characterized by its maximum of activity in 2001 and its minimum activity in 2008. Despite the decrease of the solar activity during the observations of C/2001 V1 (NEAT) and 8P/Tuttle, no change in the intensities and widths of lines were noticed. The solar activity does not appear to have any influence on the results of our work.

\section{Conclusion}

From January 2003 to September 2011, 12 comets of various origins were observed at different heliocentric distances and 48 high resolution spectra were obtained with the UVES spectrograph at VLT (ESO). Within this whole sample, we observed the three [OI] forbidden oxygen lines with high resolution and signal-to-noise and we measured the intensity ratios ($I_{6300}/I_{6364}$ and G/R~=~$I_{5577}$/($I_{6300}+I_{6364}$)) as well as the FWHM (v) of the lines. 
The results can be summarized as follows : 
\begin{enumerate}
\item The $I_{6300}/I_{6364}$ ratio (3.11~$\pm$~0.10) is in very good agreement with the branching ratio obtained by quantum mechanics, especially the work done by \cite{Galavis1997}. However, some systematic errors of the order of 0.1 are possible so that we cannot exclude the new value of \cite{Storey2000}.
\item From  theoretical values given in Table~\ref{bhardwaj}, the G/R ratio for comets observed below 2~AU (0.09~$\pm$~0.02) confirms H$_{2}$O as the main parent molecule photo-dissociating to produce oxygen atoms. However, when the comet is located at larger heliocentric distances ($r$~$\ge$~2.5~AU), the ratio increases rapidly showing that an other parent molecule is contributing. We have shown that CO$_{2}$ is the best candidate. Measuring the G/R ratios could then be a new way to estimate the abundances of CO$_{2}$, a very difficult task from the ground. Assuming that only the photo-dissociation of H$_{2}$O and CO$_{2}$ produce [OI], we found a relation between CO$_{2}$ abundance and the heliocentric distance of the comet. The C/2001 Q4 (NEAT) abundance of CO$_{2}$ at 3.7~AU is found to be $\sim$75$\%$. 
\item The intrinsic green line width is wider than the red ones by about 1 km~s$^{-1}$. Theoretical estimations considering Ly-$\alpha$ as the only excitation source for the two states lead to the conclusion that the excess energy for O($^{1}$D) is larger than for O($^{1}$S) which is in contradiction with our observations. This discrepancy might be explained by a different nature of the excitation source and/or a contribution of CO$_{2}$ as parent molecule to the O($^{1}$S) state. Indeed, \cite{Bhardwaj2012} have shown that for the photo-dissociation of CO$_{2}$, the main excitation source might rather be the 995-1165~$\r{A}$ band. To check this hypothesis quantitatively, it would be necessary to estimate theoretically the excess energy for the oxygen atoms when the wavelength band is 995-1165~\r{A}, accounting for the photo-dissociation of both H$_{2}$O and CO$_{2}$. The widths of the three [OI] lines are similar in C/2001 Q4 at $\sim$~3.7~AU. This could be in agreement with CO$_{2}$ being the main contributor for the three [OI] lines at large heliocentric distance. Other comets at large $r$ have to be observed to test this hypothesis.
\item More CO$_{2}$ (and CO) abundance determinations, together with G/R oxygen ratios and line widths at different heliocentric distances, are clearly needed in order to give a general conclusion about the oxygen parent molecule.
\item The CO$_{2}$ rich comet 103P/Hartley~2 ($\sim20\%$) does not present a high G/R ratio normally expected from Eq.~\ref{abundance}. We suggest a new value of $0.5\%$ for $\alpha^{green}_{H_{2}O}$ that could also explain the low values of G/R obtained for comets 9P/Tempel~1 and 8P/Tuttle.
\end{enumerate}

\begin{acknowledgements}
A.D thanks the support of the Belgian National Science Foundation F.R.I.A., Fonds pour la formation \`a la Recherche dans l'Industrie et l'Agriculture. E.J. is Research Associate FNRS, J.M. is Research Director FNRS and D.H. is Senior Research Associate FNRS. \\
C. Arpigny is acknowledged for the helpful discussions and constructive comments.
\end{acknowledgements}

\bibliographystyle{aa}
\bibliography{article_adecock}
\end{document}